\documentclass[preprint,12pt]{elsarticle}

\usepackage{afterpage}
\usepackage[hidelinks]{hyperref}
\usepackage{graphicx}
\usepackage[font=scriptsize]{subcaption}
\usepackage{amssymb}
\usepackage{amsmath}
\usepackage{multirow}
\usepackage[utf8]{inputenc}
\usepackage{booktabs}
\usepackage{algorithm}
\usepackage[noend]{algpseudocode}

\usepackage{xcolor}
\usepackage{xargs}
\usepackage[colorinlistoftodos,textsize=footnotesize]{todonotes}

\newcommandx{\unsure}[2][1=]{\todo[linecolor=red,backgroundcolor=red!25,bordercolor=red,#1]{#2}}
\newcommandx{\change}[2][1=]{\todo[linecolor=blue,backgroundcolor=blue!25,bordercolor=blue,#1]{#2}}
\newcommandx{\info}[2][1=]{\todo[linecolor=OliveGreen,backgroundcolor=OliveGreen!25,bordercolor=OliveGreen,#1]{#2}}

\newcommand{\ten}[1]{\ensuremath{\mathbf{#1}}}

\usepackage{lineno}

\journal{}

\begin{document}

\begin{frontmatter}

  \title{Dual-Time Smoothed Particle Hydrodynamics for Incompressible Fluid
    Simulation}
  \author[IITB]{Prabhu Ramachandran\corref{cor1}}
  \ead{prabhu@aero.iitb.ac.in}
  \author[IITB]{Abhinav Muta}
  \ead{abhinavm@aero.iitb.ac.in}
  \author[IITM]{M Ramakrishna}
  \ead{krishna@ae.iitm.ac.in}
\address[IITB]{Department of Aerospace Engineering, Indian Institute of
  Technology Bombay, Powai, Mumbai 400076}
\address[IITM]{Department of Aerospace Engineering, Indian Institute of
  Technology Madras, Chennai, 600036}

\cortext[cor1]{Corresponding author}

\begin{abstract}
  In this paper we propose a dual-time stepping scheme for the Smoothed
  Particle Hydrodynamics (SPH) method. Dual-time stepping has been used in the
  context of other numerical methods for the simulation of incompressible
  fluid flows. Here we provide a scheme that combines the entropically damped
  artificial compressibility (EDAC) along with dual-time stepping. The method
  is accurate, robust, and demonstrates up to seven times better performance
  than the standard weakly-compressible formulation. We demonstrate several
  benchmarks showing the applicability of the scheme. In addition, we provide
  a completely open source implementation and a reproducible manuscript.
  \end{abstract}


\end{frontmatter}


\section{Introduction}
\label{sec:intro}

The Smoothed Particle Hydrodynamics (SPH) method originated with the work of
\citet{monaghan-gingold-stars-mnras-77}, and \citet{lucy77} as a method to
simulate astrophysical problems. The method is grid-free and Lagrangian in
nature. It has since become a very general purpose technique and applied to a
variety of problems including incompressible fluid
flow\cite{sph:fsf:monaghan-jcp94,sph:psph:cummins-rudman:jcp:1999,isph:shao:lo:awr:2003},
solid mechanics\cite{sph-elastic-gray-jjm-cmame-2001}, and fluid-structure
interaction~\cite{rafiee:fsi-2009,khayyer-fsi-2018}.

There are several SPH schemes for simulating incompressible and weakly
compressible fluid flows. The original weakly compressible SPH scheme (WCSPH)
was proposed by \citet{sph:fsf:monaghan-jcp94}. This scheme treats the fluid
as weakly compressible with an artificial sound speed and a stiff equation of
state. This allows the scheme to utilize a hyperbolic system of equations and
integrate them in time. There are many significant variants of this scheme
including a Transport Velocity Formulation (TVF)~\cite{Adami2013} which
introduces a transport velocity to ensure particle homogeneity. The original
WCSPH and their derivatives generally suffer from a large amount of pressure
oscillations and the $\delta$-SPH
scheme~\cite{antuono-deltasph:cpc:2010,marrone-deltasph:cmame:2011} reduces
these oscillations by introducing a dissipation into the continuity equation.
Similarly, an Entropically Damped Artificial Compressibility SPH scheme
(EDAC-SPH)\cite{edac-sph:cf:2019} has been proposed which introduces entropy
by diffusing the pressure. This approach is quite similar to the $\delta$-SPH
scheme and both schemes produce superior pressure distributions. All of these
schemes employ an artificial sound speed and this places severe time step
limitations due to stability considerations.

\citet{sph:psph:cummins-rudman:jcp:1999} proposed a family of projection based
schemes for incompressible fluids. \citet{isph:shao:lo:awr:2003} and
\citet{isph:hu-adams:jcp:2007} proposed incompressible SPH (ISPH) schemes
which satisfy incompressibility by solving a pressure-Poisson equation. These
approaches eliminate the need for evolving the pressure at the sound speed and
this significantly increases the allowed time steps. The difficulty with the
projection and incompressible schemes is the requirement to solve a linear
system of equations which can be time consuming and involved. Recently, a
Predictive-Corrective ISPH (PCISPH)~\cite{pcisph:acm_tog:2009} has been
proposed for use in the graphics community for rapid simulation of
incompressible fluids. A more accurate and efficient scheme has been proposed
called the Implicit-Incompressible SPH (IISPH) \cite{iisph:ihmsen:tvcg-2014}.
The IISPH is matrix-free, and very efficient. It has been shown to be close to
an order of magnitude faster than traditional schemes. However, the IISPH can
be more involved to implement than many of the traditional WCSPH-based
schemes. Recently, another fast and matrix-free implementation of the ISPH
method has been proposed~\cite{sisph}.

In this paper we propose a new scheme for weakly-compressible fluid flows. Our
paper takes inspiration from the Artificial Compressiblility-based
Incompressible SPH (ACISPH) scheme proposed by \citet{sph:acisph:cpc:2017}.
Our scheme uses a different formulation that is also very efficient. The
original scheme was not noted in particular for its efficiency. We propose an
original derivation and suggest many improvements that make the proposed
scheme efficient. The performance is significantly better than that of the
traditional WCSPH schemes and comparable to that of the ISPH schemes without
sacrificing any accuracy or being unduly hard to implement. Our approach
employs the classic artificial compressibility of
\citet{art-compr-chorin-jcp-67} in a dual-time stepped framework. We call the
resulting scheme, DTSPH for Dual-Time stepped SPH.

\citet{fatehi-2019} propose density-based dual-timestepping schemes for the
SPH method. They propose two different formulations in order to update the
pressure in pseudo-time. They perform an accurate discretization of the
derivatives. While their method is in principle similar to the present work,
it does not demonstrate any significant performance advantage over the
traditional WCSPH schemes. Moreover, they do not demonstrate their method with
any free-surface problems. Our proposed implementation and method is however
significantly faster than the WCSPH scheme and has been demonstrated to work
well for free-surface and three-dimensional problems.
\citet{zhang-dual-time2020} on the other hand propose a dual-criteria
timestepping strategy that is quite different from the method proposed here or
in \cite{fatehi-2019,sph:acisph:cpc:2017}. They use an ``acoustic timestep''
to relax the pressure and an ``advective timestep'' to update the verlet lists
used for neighbor computation. This optimization results in a performance
improvement without significant complexity albeit at the cost of increased
memory. The present method demonstrates much improved performance and does not
involve a significant memory penalty.

The new scheme is designed to be implemented as an extension of the classic
weakly-compressible schemes. The advantage of this is that these are
relatively easy to implement, boundary conditions may be easily enforced, and
there are several well-established schemes that may be used. Although we have
not done so in this manuscript, it is important to note that this approach may
also be employed for solid mechanics problems where the artificial speed of
sound is usually very large.

The new scheme can be adapted to any WCSPH formulation which uses a density or
pressure evolution equation based on a continuity equation. We demonstrate the
scheme with the EDAC scheme~\cite{edac-sph:cf:2019}. We note that it can be
easily applied to other schemes like the $\delta$-SPH. We show how our scheme
can be used to obtain steady state solutions, although this is not particular
to the new scheme and can be easily performed for a variety of other schemes.
Obtaining steady state solutions in the context of SPH simulations is useful
in different contexts. For example, in the case of the impulsively started
flow past a complex geometry, an initial potential flow solution is useful and
this may be easily obtained using this approach.

In this manuscript we provide a new formulation as compared to the work of
\cite{sph:acisph:cpc:2017}, explore several important details for the
implementation of the scheme, and more importantly provide a high-performance,
open source implementation of the scheme. Our implementation uses the open
source PySPH framework~\cite{pysph2020, PR:pysph:scipy16} and all the code
related to the manuscript is available at
\url{https://gitlab.com/pypr/dtsph}. In order to facilitate reproducible
research, this entire manuscript is completely reproducible and every figure
in this paper is automatically generated using
\texttt{automan}~\cite{pr:automan:2018}.

In the next section we discuss the proposed DTSPH scheme in a general setting
and then discuss the SPH discretization. We briefly discuss the stability
requirements of the scheme. We show how the resulting scheme is efficient and
then proceed to simulate various standard benchmark problems. We perform
comparisons with the TVF~\cite{Adami2013}, the $\delta$-SPH
scheme~\cite{antuono-deltasph:cpc:2010}, and the EDAC
scheme~\cite{edac-sph:cf:2019} where relevant to demonstrate the accuracy and
efficiency of the proposed scheme.

\section{The dual-time SPH method}
\label{sec:dtsph}

In dual-time stepping schemes, a new time dimension called the ``dual-time'',
denoted by the variable $\tau$, is introduced. We have the following
important considerations to keep in mind. If $\ten{r}$ is the position vector
of a particle, then the \emph{real} velocity of the particle is defined as,
$\ten{V} = \frac{d \ten{r}}{dt}$. On the other hand, if the particle were to
move in pseudo-time, we define the velocity in pseudo-time as $\ten{\tilde{V}}
= \frac{d \ten{r}}{d \tau}$.

If we consider a property of the particle, $p$, then we can define a material
derivative in pseudo-time as,
\begin{equation}
  \label{eq:p_material_pseudo}
  \frac{d p}{d \tau} = \frac{\partial p}{\partial \tau} +
  \ten{\tilde{V}} \cdot \text{grad}(p).
\end{equation}

The ACISPH formulation~\cite{sph:acisph:cpc:2017} uses a non-dimensionalized
form of the equation,
\begin{equation}
  \label{eq:CE-nd}
  \frac{1}{\beta^2} \frac{\partial p^*}{\partial \tau^*} + \text{div}(\ten{V^*}) = 0.
\end{equation}
Here we use a star for the non-dimensional terms. By assuming that $p = (\rho-
\rho_0) c^2$ where $c$ is an artificial sound speed and $\rho_0$ is a
reference density, the above can be written in a dimensional form as,
\begin{equation}
  \label{eq:CE}
  \frac{\partial p}{\partial \tau} = -\rho c^2 \text{div}(\ten{V}).
\end{equation}
By rewriting the non-dimensional form in equation~\eqref{eq:CE-nd} with
suitable dimensional quantities, it is easy to see that $\beta = 1/M$ where $M
= V_{ref}/c$ is the Mach number of the reference speed of the flow, $V_{ref}$.

If we move the particles in pseudo-time, we introduce a material derivative to
get,
\begin{equation}
  \label{eq:CE_material}
  \frac{d p}{d \tau} = \ten{\tilde{V}} \cdot \text{grad}(p)
  - \rho c^2 \text{div}(\ten{V}).
\end{equation}

With the EDAC formulation~\cite{edac-sph:cf:2019}, the pressure evolution
equation also includes a diffusive term and we get,
\begin{equation}
  \label{eq:CE-weak-p-pseudo-edac}
  \frac{d p}{d \tau} = \ten{\tilde{V}} \cdot \text{grad}(p) -\rho c^2 \text{div}(\ten{V}) + \nu_{e} \nabla^2 p,
\end{equation}
where $\nu_{e}$ is an artificial viscosity parameter which is typically chosen as,
\begin{equation}
  \label{eq:nu_edac}
  \nu_{e} = \frac{h c}{16}.
\end{equation}
Here $h$ is the SPH kernel radius. In the above, we have assumed that the
particles move in pseudo-time. In later sections, we show that it is
computationally efficient to not move the particles in pseudo-time. This
results in a simpler equation for the pressure evolution without a time
derivative as,
\begin{equation}
  \label{eq:CE-p-pseudo-edac}
  \frac{\partial p}{\partial \tau} = -\rho c^2
  \text{div}(\ten{V}) + \nu_{e} \nabla^2 p.
\end{equation}

The momentum equation can be written similarly by adding a time derivative of
velocity in pseudo-time,
\begin{equation}
  \label{eq:ME-simple}
  \frac{\partial \ten{V}}{\partial \tau}
  +  \frac{d \ten{V}}{d t} =
  -\frac{1}{\rho} \text{grad}(p) + \nu \nabla^2 \ten{V} +
  \ten{f},
\end{equation}
where $\nu$ is the kinematic viscosity of the fluid and $\ten{f}$ is the body
force. Again, if we choose to move the particles in pseudo-time, we can write
this in terms of a material derivative in pseudo-time as,
\begin{equation}
  \label{eq:ME}
  \frac{d \ten{V}}{d \tau}
  +  \frac{d \ten{V}}{d t} =
  \ten{\tilde{V}} \cdot \text{grad}(\ten{V})
  -\frac{1}{\rho} \text{grad}(p) + \nu \nabla^2 \ten{V} + \ten{f}.
\end{equation}
Note the key difference here from what is proposed in
\cite{sph:acisph:cpc:2017} is that they have used $\ten{V}$ where they should
have only used $\ten{\tilde{V}}$.

There are two possible approaches we can choose for implementation:
\begin{itemize}
\item Move particles in real time and pseudo-time and use the equations
  \eqref{eq:CE_material} and \eqref{eq:ME}. Note that when using the EDAC
  formulation we would add the pressure diffusion term as shown in the right
  hand side of \eqref{eq:CE-weak-p-pseudo-edac}.
\item Move the particles only in real time and use equations \eqref{eq:CE} (or
  \eqref{eq:CE-p-pseudo-edac} in the case of the EDAC scheme) and
  \eqref{eq:ME-simple}.
\end{itemize}
We discuss these two approaches next.

\subsection{Moving particles in pseudo-time}
\label{sec:time-integration}

In this section we show how the above equations are integrated in real time
and pseudo-time. The following equations apply to each particle, $i$. We
suppress the subscript $i$ in the following to simplify the notation. We use
the index $k$ to denote pseudo-time iterations and $n$ for the real time.
Before iterating in pseudo time the particles are updated to a guessed new
state ($k=0$) for the next real time ($n+1$) using,
\begin{equation}
  \label{eq:v0}
  \ten{V}^{k=0} = \ten{V}^{n} + \Delta t {\left( \frac{d \ten{V}}{dt}
  \right)}^{n},
\end{equation}
\begin{equation}
  \label{eq:r0}
  \ten{r}^{k=0} = \ten{r}^{n} + \Delta t \ten{V}^{n} + \frac{\Delta
  t^2}{2} {\left( \frac{d \ten{V}}{dt} \right)}^{n},
\end{equation}
where ${(\frac{d \ten{V}}{dt})}^{n}$ is given by the momentum
equation~\eqref{eq:ME-simple} without the partial derivative of velocity in
pseudo time (i.e.\ considering $\frac{\partial \ten{V}}{\partial \tau} = 0$).
Then the integration in pseudo time proceeds in the following fashion, with
$\ten{\tilde{V}}^{k=0} = 0 $ as the starting value,

\begin{equation}
  \label{eq:rkstar}
  \ten{r}^{k+1/2} = \ten{r}^k + \frac{1}{2}\Delta \tau \ten{\tilde{V}}^k,
\end{equation}
\begin{equation}
  \label{eq:pkstar}
  p^{k+1/2} = p^k + \frac{1}{2} \Delta \tau {\left(  \frac{d p}{d \tau}\right)}^k,
\end{equation}

\begin{equation}
  \label{eq:vkp1}
  \ten{V}^{k+1} = \ten{V}^k +
  \Delta \tau {\left( \frac{d \ten{V}}{d \tau} \right)}^{k+1/2},
\end{equation}
\begin{equation}
  \label{eq:rkp1}
  \ten{r}^{k+1} = \ten{r}^{k+1/2} + \frac{1}{2}\Delta \tau \ten{\tilde{V}}^{k+1},
\end{equation}
\begin{equation}
  \label{eq:pkp1}
  p^{k+1} = p^{k+1/2} +
  \frac{1}{2} \Delta \tau {\left(\frac{d p}{d \tau}\right)}^{k+1/2}.
\end{equation}
In addition to these we have,
\begin{equation}
  \label{eq:vtilde}
  \ten{\tilde{V}}^{k+1} = {\left( \frac{d \ten{V}}{d \tau} \right)}^{k+1/2}
  \Delta t.
\end{equation}
This is to ensure consistency of the motion. A detailed proof for
equation~\eqref{eq:vtilde} is provided in the appendix at the end of this
manuscript.

We need an expression for the term $\frac{d \ten{V}}{dt}$ in the momentum
equation~\eqref{eq:ME}. We obtain this using an implicit three-point backward
difference scheme to discretize the real-time derivative,
\begin{equation}
  \label{eq:fd-dvdt-real}
  {\left( \frac{d \ten{V}}{d t} \right)}^{n+1}
  = \frac{3\ten{V}^{n+1} - 4 \ten{V}^{n} + \ten{V}^{n-1}}{2 \Delta t}
  + {O(\Delta t)}^2.
\end{equation}
Substitute $\ten{V}^{k+1}$ instead of $\ten{V}^{n+1}$
in~\eqref{eq:fd-dvdt-real} and add and subtract the term $3\ten{V}^{k}$ in the
numerator and rearrange terms to get,
\begin{equation}
  \label{eq:fd-dvdt}
  {\left( \frac{d \ten{V}}{d t} \right)}^{n+1}
  = \frac{3 \Delta \tau}{2 \Delta t}
  {\left( \frac{d \ten{V}}{d \tau} \right)}^{k+1/2}
  + \frac{3\ten{V}^{k} - 4 \ten{V}^{n} + \ten{V}^{n-1}}{2 \Delta t}
  + {O(\Delta t)}^2.
\end{equation}
If we use equation~\eqref{eq:ME}, we can rewrite the above as,
\begin{equation}
  \label{eq:dv_dtau_exact}
  \begin{split}
    \left(  \frac{d \ten{V}}{d \tau}\right)^{k+1/2} \approx &
    \left \{
      \ten{\tilde{V}} \cdot \text{grad}(\ten{V})
      -\frac{1}{\rho} \text{grad}(p) + \nu \nabla^2 \ten{V} + \ten{f} \right. \\
    & \left. - \frac{(3\ten{V}^{k} - 4 \ten{V}^{n} + \ten{V}^{n-1})}{2 \Delta t}
    \right \} \left( \frac{2 \Delta t}{2\Delta t + 3 \Delta \tau} \right).
\end{split}
\end{equation}
We may now discretize the right hand side using SPH and find the acceleration
to the velocity in pseudo-time. This is done in section~\ref{sec:sph}.

\subsection{Fixed particles in pseudo-time}
\label{sec:time-integration-frozen}

In this section we consider the case where the particles are not moved in
pseudo-time. In this case, we can perform the integration as follows,
\begin{equation}
  \label{eq:pkstar_frozen}
  p^{k+1/2} = p^k + \frac{1}{2} \Delta \tau {
    \left(  \frac{\partial p}{\partial \tau}\right)}^k,
\end{equation}

\begin{equation}
  \label{eq:vkp1_frozen}
  \ten{V}^{k+1} = \ten{V}^k +
  \Delta \tau {\left( \frac{\partial \ten{V}}{\partial \tau} \right)}^{k+1/2},
\end{equation}
\begin{equation}
  \label{eq:pkp1_frozen}
  p^{k+1} = p^{k+1/2} +
  \frac{1}{2} \Delta \tau {\left(\frac{\partial p}{\partial \tau}\right)}^{k+1/2}.
\end{equation}
Further, starting from equation~\eqref{eq:vkp1_frozen}, the equation
corresponding to \eqref{eq:fd-dvdt} becomes,
\begin{equation}
  \label{eq:fd-dvdt_frozen}
  {\left( \frac{d \ten{V}}{d t} \right)}^{n+1}
  = \frac{3 \Delta \tau}{2 \Delta t}
  {\left( \frac{\partial \ten{V}}{\partial \tau} \right)}^{k+1/2}
  + \frac{3\ten{V}^{k} - 4 \ten{V}^{n} + \ten{V}^{n-1}}{2 \Delta t},
\end{equation}
where, the material derivative in the first term on the right hand side is
replaced with a partial pseudo-time derivative. The
equation~\eqref{eq:dv_dtau_exact} also changes appropriately to no longer
include the pseudo-time advective term, resulting in
\begin{equation}
  \label{eq:dv_dtau_exact}
  \begin{split}
    \left(  \frac{\partial \ten{V}}{\partial \tau}\right)^{k+1/2} \approx &
    \left \{
      -\frac{1}{\rho} \text{grad}(p) + \nu \nabla^2 \ten{V} + \ten{f} \right. \\
    & \left. - \frac{(3\ten{V}^{k} - 4 \ten{V}^{n} + \ten{V}^{n-1})}{2 \Delta t}
    \right \} \left( \frac{2 \Delta t}{2\Delta t + 3 \Delta \tau} \right).
\end{split}
\end{equation}
We note that usually the velocity of a particle in pseudo-time,
$\ten{\tilde{V}}$, is very small and this makes the changes to the position
even smaller. This means that it is computationally efficient to fix the
position of the particles in pseudo-time. In addition, even if we were to move
the particles, we do not need to recompute the neighbor information.

Once the pseudo-time iterations are completed, we update the particle
positions using,
\begin{equation}
  \label{eq:r-update}
  \ten{r}^{n+1}= \ten{r}^n + \frac{\Delta t}{2} \left( \ten{V}^n  + \ten{V}^{n+1} \right).
\end{equation}

\subsection{Steady state solutions}
\label{sec:steady-state}

We can use the dual-time to seek a solution to steady state problems. To do
this we set the partial derivative in time to zero and retain only the pseudo
time derivative. Further, we do not move the particles at all.  We start with
the following form of the continuity equation,
\begin{equation}
  \label{eq:CE-steady}
  \frac{\partial p}{\partial \tau} + \frac{\partial p}{\partial t} +
  \ten{V} \cdot \text{grad}(\ten{V}) =
  - \rho c^2 \text{div}(\ten{V}).
\end{equation}
We set the partial derivative in time to zero and this results in using the
following form of equation for the evolution of the pressure,
\begin{equation}
  \label{eq:CE-steady}
  \frac{\partial p}{\partial \tau} =
  -\ten{V} \cdot \text{grad}(\ten{V})
  - \rho c^2 \text{div}(\ten{V}).
\end{equation}
We note that we have not used the EDAC scheme in this case.
Similarly, the equation~\eqref{eq:ME-simple} for momentum reduces to,
\begin{equation}
  \label{eq:ME-steady}
  \frac{\partial \ten{V}}{\partial \tau} =
  - \ten{V} \cdot \text{grad}(\ten{V})
  - \frac{1}{\rho} \text{grad}(p) +
  \nu \nabla^2 \ten{V} +
  \ten{f}.
\end{equation}
Here, we have moved the convection term to the right side and removed any
partial time derivatives. This can be easily solved purely in pseudo-time
while keeping the particles fixed in space. Technically, we could replace
$\tau$ with $t$. However, the dual-time offers a convenient perspective for
seeking a steady state solution iteratively.

The above approach to obtain steady state solutions is simple and not tied to
any particular SPH scheme. Any scheme that uses a density or pressure
evolution equation that is dependent on the divergence of the velocity will
work. The approach is numerically efficient as it does not require any
re-computation of neighbors. While simulating the steady-state problem, we
iterate until the changes in pseudo-time become small enough.

\section{SPH discretization}
\label{sec:sph}

The basic scheme discussed in the previous section should work for any
particular SPH discretization of the momentum and pressure equations. In the
following, we use a WCSPH formulation for the SPH discretization. We keep
density fixed as per the original problem. We consider two different cases one
where we move the particles in pseudo-time and the other where the particles
are fixed in pseudo-time.

\subsection{Moving particles in pseudo-time}
\label{sec:sph-moving}

When the particles move in pseudo-time, the pressure evolution is computed
using equation~\eqref{eq:CE-weak-p-pseudo-edac}.  This is discretized as,
\begin{equation}
  \label{eq:dp-dtau}
  \begin{split}
  \frac{d p_i }{d \tau} = & \ten{\tilde{V}}_i \cdot \sum_{j\in N(i)}
  m_j \left( \frac{p_i}{\rho_i^2} + \frac{p_j}{\rho_j^2}  \right)
  \nabla W_{ij}  \\
  + &
  \sum_{j\in N(i)} \frac{m_j \rho_i}{\rho_j} c^2 \
    \ten{V}_{ij} \cdot \nabla W_{ij}  \\
  + & \sum_{j\in N(i)} \frac{4 m_j}{ (\rho_i + \rho_j)\rho_j} \nu_e \frac{p_{ij}} {(\ten{r}_{ij}^2
    + \ten{h}_{ij}^2)}  \nabla W_{ij} \cdot \ten{x}_{ij},
\end{split}
\end{equation}
where $i$ denotes the $i$'th particle, $\ten{V}_{ij} = \ten{V}_i - \ten{V}_j$,
$p_{ij} = p_i - p_j$, $\ten{r}_{ij} = \ten{r}_i - \ten{r}_j$, $\ten{x}_{ij} =
\ten{x}_i - \ten{x}_j$, $W_{ij} = W(|\ten{r_i} - \ten{r_j}|, h)$, is the SPH
smoothing kernel with $h$ as the smoothing length of the kernel and $\nu_e$ is
the artificial viscosity coefficient used for the EDAC
scheme~\cite{edac-sph:cf:2019} as shown in equation~\eqref{eq:nu_edac}. The
kernel is compact so the summation is over all the nearest neighbor particles
that influence the particle $i$, $N(i)$. We note that the second term in the
right hand side of equation~\eqref{eq:CE-weak-p-pseudo-edac} corresponds to a
density diffusion term employed in the $\delta$-SPH
scheme\cite{antuono-deltasph:cpc:2010}.

For the momentum equation where the particles move in pseudo-time
given in \eqref{eq:ME}, we can write,
\begin{equation}
  \label{eq:mom-sph}
  \begin{split}
    \frac{d \ten{V}_i}{d \tau} + \frac{d \ten{V}_i}{d t} =
    & \ten{\tilde{V}_i} \cdot \sum_{j \in N(i)} \left(\ten{V}_i \frac{m_i}{\rho_i}
    + \ten{V}_j \frac{m_j}{\rho_j}\right) \nabla W_{ij}  \\
  & - \sum_{j\in N(i)} m_j \left(\frac{p_i}{\rho_i^2} + \frac{p_j}{\rho_j^2}
  + \Pi_{ij} \right)
    \nabla W_{ij} \\
    & + \sum_{j\in N(i)} m_j \frac{4 \nu \nabla W_{ij}\cdot
      \ten{r}_{ij}}{(\rho_i + \rho_j) (r_{ij}^2 + \eta h_{ij}^2)} \ten{V}_{ij}
    + \ten{g}_i,
  \end{split}
\end{equation}
where $\Pi_{ij}$ is the artificial viscosity term~\cite{monaghan-review:2005}
added to the momentum equation is given by,
\begin{align}
  \label{eq:mom-av}
  \Pi_{ij} =
  \begin{cases}
\frac{-\alpha h_{ij} \bar{c}_{ij} \phi_{ij}}{\bar{\rho}_{ij}}
  & \ten{v}_{ij}\cdot \ten{r}_{ij} < 0, \\
  0 & \ten{v}_{ij}\cdot \ten{r}_{ij} \ge 0, \\
\end{cases}
\end{align}
\begin{equation}
  \label{eq:av-phiij}
  \phi_{ij} = \frac{\ten{v}_{ij} \cdot \ten{r}_{ij}}{r^2_{ij} + \epsilon h^2_{ij}},
\end{equation}
where $\ten{r}_{ij} = \ten{r}_i - \ten{r}_j$, $\ten{v}_{ij} = \ten{v}_i -
\ten{v}_j$, $h_{ij} = (h_i + h_j)/2$, $\bar{\rho}_{ij} = (\rho_i + \rho_j)/2$,
$\bar{c}_{ij} = (c_i + c_j) / 2$, $\eta=0.01$, and $\alpha$ is the artificial
viscosity parameter. Note that in the above equations the viscosity is modeled
using the discretization proposed by \citet{morris-lowRe-97}.

We now write out the final form of the rate of change of the velocity in
pseudo-time for the case where the particles are moving which we get by
substituting equation~\eqref{eq:fd-dvdt} in the momentum
equation~\eqref{eq:ME}, and discretizing the equations using the SPH
formulation to get,
\begin{equation}
\label{eq:discrete_ME}
\begin{split}
    \frac{d \ten{V}_i}{d \tau} =
    & \left\{\ten{\tilde{V}_i} \cdot \sum_{j\in N(i)} \left(\ten{V}_i \frac{m_i}{\rho_i}
    + \ten{V}_j \frac{m_j}{\rho_j}\right) \nabla W_{ij} \right. \\
  & -\sum_{j\in N(i)} m_j \left(\frac{p_i}{\rho_i^2} + \frac{p_j}{\rho_j^2}
  + \Pi_{ij} \right)
    \nabla W_{ij}\\
    & + \sum_{j\in N(i)} m_j \frac{4 \nu \nabla W_{ij}\cdot
      \ten{r}_{ij}}{(\rho_i + \rho_j) (r_{ij}^2 + \eta h_{ij}^2)} \ten{V}_{ij}
    + \ten{g}_i \\
    & \left. - \frac{(3\ten{V}_i^{k} - 4\ten{V}_i^{n} + \ten{V}_i^{n-1})}{2 \Delta t}
    \right \} \left( \frac{2 \Delta t}{2 \Delta t + 3 \Delta \tau} \right) .
\end{split}
\end{equation}
It is important to note here that while we have used standard WCSPH
discretizations, the DTSPH formulation would work just as well with any other
SPH discretization that is based on a weakly-compressible formulation. While
in principle this may be extended to Riemann-solver based SPH
formulations~\cite{ferrari-riemann-sph-2009}, this is not explored in the
present work.

We observe that in equation~\eqref{eq:discrete_ME}, we require the velocity at
the current time and the previous time. When starting the simulations, if we
do not have an exact solution, we assume that $\ten{V}^{-1} = \ten{V}^0$.

\subsection{Fixed particles in pseudo-time}
\label{sec:sph-fixed}

In the case the particles are fixed in pseudo-time. The right hand side of the
equation~\eqref{eq:CE} is discretized as,
\begin{equation}
  \label{eq:dp-dtau-frozen}
  \frac{\partial p_i }{\partial \tau} =
  \sum_{j\in N(i)} \frac{m_j \rho_i}{\rho_j} c^2 \ \ten{V}_{ij} \cdot \nabla W_{ij}.
\end{equation}
In the case of the EDAC scheme, the right hand side of the
equation~\eqref{eq:CE-p-pseudo-edac} is discretized as,
\begin{equation}
  \label{eq:dp-dtau-frozen-edac}
  \frac{\partial p_i }{\partial \tau} =
  \sum_{j\in N(i)} \frac{m_j \rho_i}{\rho_j} c^2 \ \ten{V}_{ij} \cdot \nabla
  W_{ij} + \sum_{j\in N(i)} \frac{4 m_j}{ (\rho_i + \rho_j)\rho_j}
  \nu_e \frac{p_{ij}} {(\ten{r}_{ij}^2
    + \ten{h}_{ij}^2)}  \nabla W_{ij} \cdot \ten{x}_{ij},
\end{equation}
For the momentum equation given in \eqref{eq:ME-simple}, we can write,
\begin{align}
  \label{eq:mom-sph-frozen}
  \begin{split}
    \frac{\partial \ten{V}_i}{\partial \tau} + \frac{d \ten{V}_i}{d t} = &
    \sum_{j\in N(i)} -m_j \left(\frac{p_i}{\rho_i^2} + \frac{p_j}{\rho_j^2}
    + \Pi_{ij} \right)
    \nabla W_{ij} \\
    & + \sum_{j\in N(i)} m_j \frac{4 \nu \nabla W_{ij}\cdot
      \ten{r}_{ij}}{(\rho_i + \rho_j) (r_{ij}^2 + \eta h_{ij}^2)} \ten{V}_{ij}
    + \ten{g}_i.
  \end{split}
\end{align}

By substituting equation~\eqref{eq:mom-sph-frozen} in
equation~\eqref{eq:fd-dvdt_frozen} or equation~\eqref{eq:mom-sph} in
equation~\eqref{eq:fd-dvdt}, we get an equation for either $\frac{\partial
  \ten{V}}{\partial \tau}$ or $\frac{d \ten{V}}{d \tau}$. We can then use this
to integrate the set of equations \eqref{eq:rkstar} -- \eqref{eq:vtilde} or
\eqref{eq:pkstar_frozen} -- \eqref{eq:pkp1_frozen}. Note that we first update
the velocity and position of the particle as per equation~\eqref{eq:v0} and
\eqref{eq:r0}.

The above equations govern the velocity and pressure of the fluids in the
simulation. In order to satisfy the boundary conditions when solids are
present we use an implementation of the boundary conditions presented in
\cite{Adami2012} where the pressure and the ghost velocity of the solid walls
are set. Furthermore, following the work of
\cite{hughes-graham:compare-wcsph:jhr:2010}, we ensure that the pressure is
always positive on the solid walls to prevent particles from sticking to them
in our free surface simulations. We do not impose any specific free-surface
boundary conditions.

\subsection{Steady state solutions}
\label{sec:steady-sph}

For the case of the steady state simulations, we do not move the particles in
time and use the original distribution of particles. This is reasonable as
steady solutions are usually sought where the geometry and boundaries are
fixed. As discussed earlier, this leads to a very efficient solution procedure.
For the pressure evolution we simply use equation~\eqref{eq:dp-dtau-frozen}.
For the momentum equation we use the following discretization for
equation~\eqref{eq:ME-steady},
\begin{equation}
  \label{eq:mom-steady}
  \begin{split}
    \frac{\partial \ten{V}_i}{\partial \tau} = &
    \sum_{j\in N(i)} -m_j \left(\frac{p_i}{\rho_i^2} + \frac{p_j}{\rho_j^2}
    + \Pi_{ij} \right)
    \nabla W_{ij}
   + \sum_{j\in N(i)} m_j \frac{4 \nu \nabla W_{ij}\cdot
      \ten{r}_{ij}}{(\rho_i + \rho_j) (r_{ij}^2 + \eta h_{ij}^2)} \ten{V}_{ij} \\
   & + \ten{g}_i
   - \ten{V}_i \cdot \sum_{j}^{N} \left( \ten{V}_i \frac{m_i}{\rho_i}
    + \ten{V}_j \frac{m_j}{\rho_j} \right) \nabla W_h,
\end{split}
\end{equation}

By solving these until the pseudo-time derivatives are small, we can obtain
steady state solutions. This can be implemented very efficiently. The
neighbors can be computed once and never need to be updated. The time step
restrictions though continue to be as per the original weakly-compressible
scheme.

\subsection{Stability and convergence}
\label{sec:stability}

It is important to choose the real and pseudo-timesteps carefully. We choose
$\Delta t$ such that $\Delta t \ten{V}_{max} = C h$, and $C$ is around 0.25.
This is similar to the timesteps used in the ISPH schemes.

We choose $1 < \beta < 20$, recall that $\beta = 1/M = c/V_{ref}$. We choose
$\Delta \tau$ as $ \Delta t / \beta$. The choice of these parameters is due to
the following observations:
\begin{itemize}
\item The real time step is limited by the amount of permitted motion of the
  particles in one time step.

\item The pseudo-time step can be seen to be essentially similar to the
  original weakly-compressible scheme and is therefore limited by the speed of
  sound. The pressure waves travel at the speed of sound and therefore the
  pseudo-timesteps should be limited to around $\Delta t/\beta$.
\end{itemize}

We use the following approach to decide when to stop iterating in pseudo-time.
The user specifies a particular tolerance, $\epsilon$, which determines the
termination of the pseudo-time iterations. During every pseudo-time iteration
we compute the mean rate of change in the pressure, let us call this quantity
$\delta p/\delta \tau$. We also compute the mean value of $|\ten{\tilde{V}}|$
for all particles. When checking for convergence we ensure the following,
\begin{align}
  \label{eq:conv}
  \begin{split}
    \frac{|\delta p/\delta\tau| \Delta t}{\rho c^2} \ &  < \ \epsilon, \\
    \frac{|\ten{\tilde{V}}|}{V_{ref}} \ & < \ \epsilon.
  \end{split}
\end{align}
Note that we multiply the rate of change of pressure by $\Delta t$ in order to
ensure that the change over several pseudo-iterations would be accounted for.

Due to the inaccuracies of the SPH approximations and particle disorder, it is
likely that the divergence does not become less than the tolerance and that
the derivatives do not reduce. In order to prevent needless iterations we keep
track of the changes in each pseudo-time iteration and stop iterations if the
peak-to-peak relative changes in the last 3 or 4 iterations is less than 5\%.
This ensures that if the pressure and velocity do not change with increasing
iterations we stop the iterations. This works very well in practice. We also
stop iterating if there are more than 1000 iterations. In practice for
reasonable tolerance values (larger than $10^{-5}$) we typically have far less
than 50 iterations per real time step. Our default tolerance is $\epsilon =
10^{-3}$.

\subsection{Particle shifting}
\label{sec:sph-shifting}
We use the shifting algorithm proposed
by~\citet{diff_smoothing_sph:lind:jcp:2009} which is based on Fickian
diffusion. All the the test cases use shifting unless stated explicitly. The
positions are shifted to their new position,
\begin{equation}%
  \label{eq:shift}
  \ten{r}_{i'} = \ten{r}_{i} + \delta \ten{r}_{ii'},
\end{equation}
where,
\begin{equation}%
  \label{eq:shift-deltar}
  \delta \ten{r}_{ii'} = - \frac{h^2}{2} \sum_{j \in N(i)} \frac{m_j}{\rho_j} \left (1 + \frac{W(\ten{r}_{ij},
    h)}{W(\Delta x, h)} \right) \nabla W_{ij},
\end{equation}
where the present implementation uses $\Delta x=h$, the particle smoothing
length. After shifting, all the fluid properties are corrected using Taylor
series approximation as follows,
\begin{equation}%
  \label{eq:shift-correct}
  \varphi_{i'} = \varphi_i + (\nabla \varphi)_i\cdot \delta \ten{r}_{ii'},
\end{equation}
where $\varphi_i$ and $\varphi_{i'}$ denotes the fluid property before and
after shifting respectively for $i^{\text{th}}$ fluid particle.

We apply the free-surface identification as suggested by
\citet{diff_smoothing_sph:lind:jcp:2009} where a particle is considered as a
free-surface particle if $\nabla \cdot \ten{r}_i < d - 1/2$ where $d$ is the
spatial dimension of the problem. Such free-surface particles are not shifted.

\subsection{Complete algorithm}
\label{sec:algorithm}

We summarize the proposed scheme using pseudo-code for clarity. The case of
particles moving in pseudo-time is shown in Algorithm~\ref{alg:moving}. The
case of particles fixed in pseudo-time is shown in Algorithm~\ref{alg:fixed}.
\begin{algorithm}[!ht]
  \caption{Moving particles in pseudo-time, see Section~\ref{sec:sph-moving}.}%
  \label{alg:moving}
  \begin{algorithmic}[1]
    \While{$t < t_{\text{final}}$}
    \For{all particles}
    \State{compute $\left(\frac{d p_i}{d \tau}\right)^{k}$ using equation~\eqref{eq:dp-dtau}}
    \State{compute $\left(\frac{d \ten{V}_i}{d \tau}\right)^{k}$ using equation~\eqref{eq:discrete_ME}}
    \EndFor%
    \For{all particles}
    \State{predict velocity using equation~\eqref{eq:v0}}
    \State{predict position using equation~\eqref{eq:r0}}
    \EndFor%
    \While{check convergence using equation~\eqref{eq:conv}}
    \For{all particles}
    \State{update $\left(\frac{d p_i}{d \tau}\right)^{k+\frac{1}{2}}$
      using equation~\eqref{eq:dp-dtau}}
    \State{update $\left(\frac{d \ten{V}_i}{d \tau}\right)^{k+\frac{1}{2}}$
      using equation~\eqref{eq:discrete_ME}}
    \EndFor%
    \For{all particles}
    \State{update $\tilde{\ten{V}}^{k+1}$ using equation~\eqref{eq:vtilde}}
    \EndFor%
    \For{all particles}
    \State{update to $\ten{r}^{k+1}$, $\ten{v}^{k+1}$, and $\ten{p}^{k+1}$ using
      equations~\eqref{eq:rkstar}-\eqref{eq:pkp1}}
    \EndFor%
    \EndWhile{}
    \State{update positions to $\ten{r}^{n+1}$ using equation~\eqref{eq:r-update}}
    \State{shift particles using equation~\eqref{eq:shift}}
    \State{correct the velocities and pressure using equation~\eqref{eq:shift-correct}}
    \EndWhile{}
  \end{algorithmic}
\end{algorithm}
\begin{algorithm}[t]
  \caption{Fixed particles in pseudo-time, see Section~\ref{sec:sph-fixed}.}%
  \label{alg:fixed}
  \begin{algorithmic}[1]
    \While{$t < t_{\text{final}}$}
    \For{all particles}
    \State{compute $\left(\frac{\partial p_i}{\partial \tau}\right)^{k}$ using
      equation~\eqref{eq:dp-dtau-frozen-edac}}
    \State{compute $\left(\frac{\partial \ten{V}_i}{\partial \tau}\right)^k$ using
      equation~\eqref{eq:mom-sph-frozen}}
    \EndFor%
    \For{all particles}
    \State{predict velocity using equation~\eqref{eq:v0}}
    \State{predict position using equation~\eqref{eq:r0}}
    \EndFor%
    \While{check convergence using equation~\eqref{eq:conv}}
    \For{all particles}
    \State{compute $\left(\frac{\partial p_i}{\partial \tau}\right)^{k+\frac{1}{2}}$
      using equation~\eqref{eq:dp-dtau-frozen-edac}}
    \State{update  $\left(\frac{\partial \ten{V}_i}{\partial \tau}\right)^{k+\frac{1}{2}}$
      using equation~\eqref{eq:mom-sph-frozen}}
    \EndFor%
    \For{all particles}
    \State{update to $\ten{v}^{k+1}$, and $\ten{p}^{k+1}$ using
      equations~\eqref{eq:pkstar_frozen}-\eqref{eq:pkp1_frozen}}
    \EndFor%
    \EndWhile{}
    \State{update positions to $\ten{r}^{n+1}$ using equation~\eqref{eq:r-update}}
    \State{shift particles using equation~\eqref{eq:shift}}
    \State{correct the velocities and pressure using equation~\eqref{eq:shift-correct}}
    \EndWhile{}
  \end{algorithmic}
\end{algorithm}

\section{Results and discussion}
\label{sec:results}

In this section we perform various numerical experiments using standard
benchmark problems. We explore the following specific questions using the
Taylor-Green problem which has an exact solution:
\begin{itemize}
\item Is it worth moving the particles in pseudo-time or can we freeze the
  particles? This has significant performance implications.
\item What possible values of $\beta$ can be used?
\item What suitable values of the tolerance $\epsilon$ can be chosen and what
  does this imply for accuracy?
\end{itemize}

Once these are explored, a suite of test problems are simulated with the DTSPH
scheme and compared with other schemes like the standard
WCSPH~\cite{sph:fsf:monaghan-jcp94}, transport velocity formulation
(TVF)~\cite{Adami2013}, Entropically damped artificial compressibility (EDAC)
~\cite{edac-sph:cf:2019}, and the $\delta$-SPH
scheme~\cite{antuono-deltasph:cpc:2010}. Except where noted, the DTSPH scheme
uses the EDAC formulation in all the simulations below i.e.\ the pressure
evolution uses equation~\eqref{eq:dp-dtau-frozen-edac}.

The TVF, EDAC, $\delta$-SPH, and WCSPH schemes are part of the
PySPH~\cite{pysph2020,PR:pysph:scipy16} framework. All the results presented below
are automated and the code for the benchmarks is available at
\url{https://gitlab.com/pypr/dtsph}. The tools used to automate the results
are described in detail in~\cite{pr:automan:2018}. This allows us to
automatically reproduce every figure and table in this manuscript.

All the simulations are performed on a four core Intel (R) Core (TM) i$5-7400$
CPU with a clock speed of 3.00GHz. The problems are executed on four cores
using OpenMP.  We use the DTSPH scheme along with the EDAC in all results
below unless explicitly mentioned otherwise.

\subsection{Taylor-Green problem}
\label{sec:tgv}

The Taylor-Green problem is a classical problem which is periodic along both
$x$ and $y$ axes and has an exact solution. This is a particularly challenging
problem for SPH~\cite{Adami2013, sph:acisph:cpc:2017, fatehi-2019} since the
particles move along the streamlines towards a stagnation point leading to
particle disorder.

The exact solution for the Taylor-Green problem is given by
\begin{align}
  \label{eq:tgv_sol}
  u &= - U e^{bt} \cos(2 \pi x) \sin(2 \pi y), \\
  v &=   U e^{bt}\sin(2 \pi x) \cos(2 \pi y), \\
  p &=  -U^2 e^{2bt} (\cos(4 \pi x) + \cos(4 \pi y))/4,
\end{align}
where $U$ is chosen as $1m/s$, $b=-8\pi^2/Re$, $Re=U L /\nu$, and $L=1m$.

The Reynolds number $Re$ is set to 100 and various cases are tested to better
understand the scheme. For all the simulations, the quintic spline kernel is
used with $h/\Delta x = 1.0$, no artificial viscosity is used. The following
cases are considered,
\begin{itemize}
\item a comparison of results when particles are either advected or frozen in
  pseudo-time;
\item the effect of changing the artificial speed of sound, $\beta$;
\item the effect of changing the convergence tolerance, $\epsilon$;
\item comparison of results with different schemes; and
\item comparison of the results with different number of particles and with
  different Reynolds numbers.
\end{itemize}
The results are compared against the exact solution. Particle plots are also
shown wherever necessary as the error plots do not always reveal any particle
disorder, particle clumping, or voiding occurring in the flow. In addition, we
compare the performance of the schemes where the difference is noticeable.

We first discuss a simple method of initializing the particles that we use to
compare all the schemes in a consistent and fair manner.
\subsubsection{Perturbation in initial positions}
\label{sub:sec:tg:perturb}

As discussed earlier, it is not always easy to obtain good results for the
Taylor-Green problem. When the particles are initially distributed in a
uniform manner, they tend to move towards the stagnation points and this often
leads to severe particle disorder. In~\cite{edac-sph:cf:2019}, it was found
that this problem can be reduced by introducing a small amount of noise in the
initial particle distribution. To this end, a small random displacement is
given to the particles with a maximum displacement of $\Delta x /10$. The
random numbers are drawn from a uniform distribution and the random seed is
kept fixed leading to the same distribution of particles for all cases with
the same number of particles. This allows us to perform a fair comparison.
Initially the particles are arranged in a $100 \times 100$ grid, with
smoothing length of $h/\Delta x = 1$, $\Delta t = 0.00125$, $\epsilon =
10^{-4}$, and simulated for $2.5 s$ with $\beta=5$.

The decay rate is computed by computing the magnitude of maximum velocity
$|\ten{V}_{\max}|$ at each time step, the $L_1$ error is computed as the
average value of the difference between the exact velocity magnitude and the
computed velocity magnitude, given as
\begin{equation}
  \label{eq:tg:l1}
  L_1 = \frac{\sum_i |\ten{V}_{i, computed} - \ten{V}_{i, exact}|}
  {\sum_i |\ten{V}_{i, exact}|},
\end{equation}
where $\ten{V}_i$ is computed at the particle positions for each particle $i$
in the flow.
\begin{figure}[!h]
  \centering
  \begin{subfigure}{0.48\textwidth}
    \centering
    \includegraphics[width=1.0\textwidth]{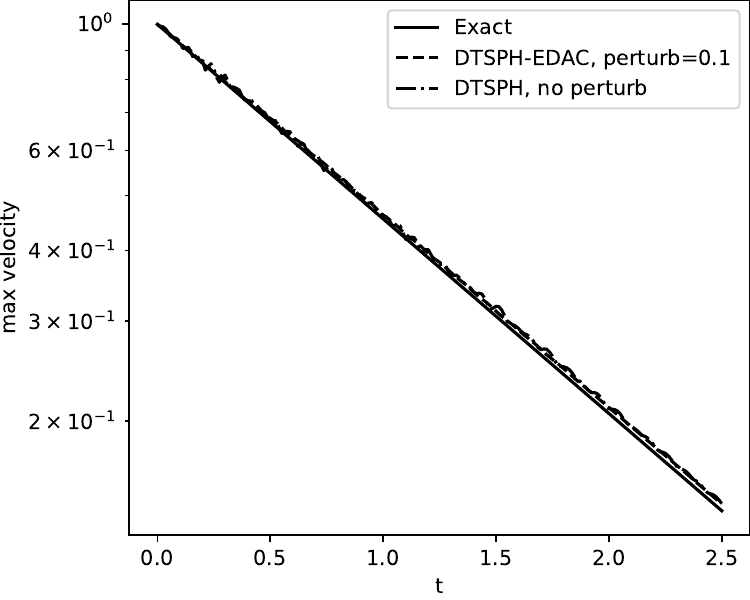}
    \subcaption{Decay of maximum velocity with time.}\label{fig:tg:pert:decay}
  \end{subfigure}
  \begin{subfigure}{0.48\textwidth}
    \centering
    \includegraphics[width=1.0\textwidth]{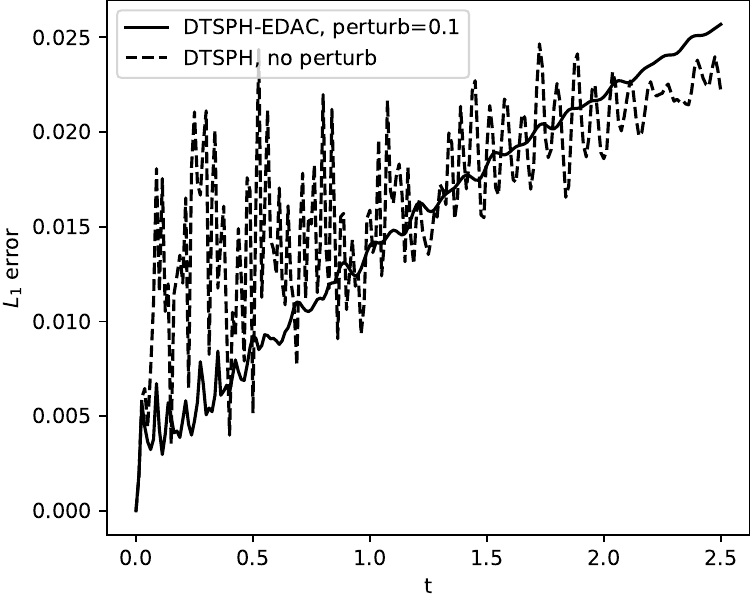}
    \subcaption{$L_1$ error of velocity magnitude with time.}\label{fig:tg:pert:l1}
  \end{subfigure}
  \caption{Comparison of perturbation (of atmost $\Delta x/10$) and without any
    perturbation.}
\label{fig:tg:pert}
\end{figure}
Fig.~\ref{fig:tg:pert:decay} shows the decay of the velocity compared with the
exact solution for the case where there is no initial perturbation and with a
small amount of perturbation (of at most $\Delta x/10$). As can be clearly
seen, the introduction of the perturbation significantly improves the results.
This is clearly seen in the $L_1$ norm of the error in the velocity magnitude
in Fig.~\ref{fig:tg:pert:l1}. While we have not shown this, the results are
similar for simulations made using most other schemes including the new
scheme, TVF, and the EDAC. Given this, we henceforth use a small initial
perturbation for the results for the Taylor-Green vortex problem.

\subsubsection{Advection of particles in pseudo-time}
\label{sub:sec:tg:advect}

As discussed earlier, it is important to study the effect of moving the
particles in pseudo-time as against keeping them frozen in pseudo-time.  If we
find that there is no significant advantage gained by moving the particles in
pseudo-time, we can simplify the implementation of the scheme as well as
improve its performance considerably.

We consider the Taylor Green problem and compare the results of simulations
where we advect the particles and where we hold them frozen. The rest of the
parameters are held fixed. The same small perturbation is added to the initial
particle position. Parameters used for this simulation are, initial particle
spacing $\Delta x = \Delta y = 0.01$, Reynolds number $Re=100$, the value of
$\beta=5$, time step $\Delta t = 0.00125$, and tolerance of $\epsilon =
10^{-6}$.
\begin{figure}[!h]
  \centering
  \begin{subfigure}[!h]{0.48\textwidth}
    \includegraphics[width=\linewidth]{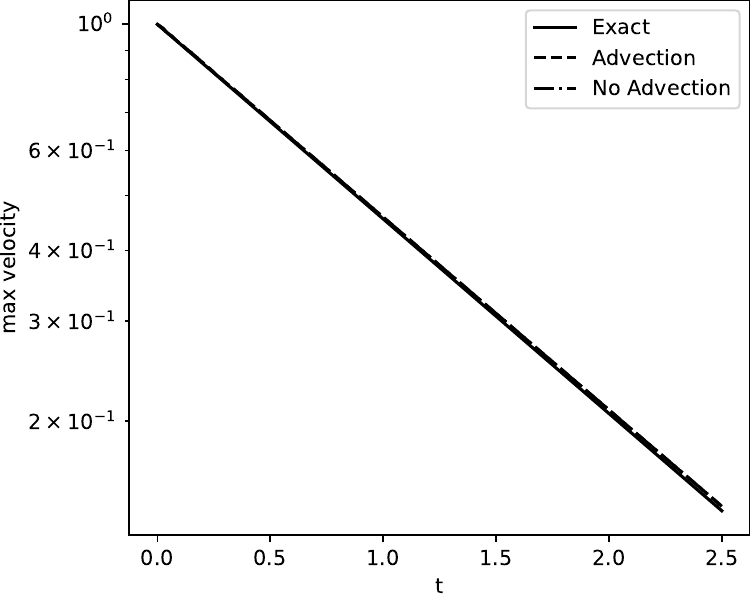}
    \subcaption{Decay of maximum velocity with time.}\label{fig:tg:advect:decay}
  \end{subfigure}
  \begin{subfigure}[!h]{0.48\textwidth}
    \centering
    \includegraphics[width=\linewidth]{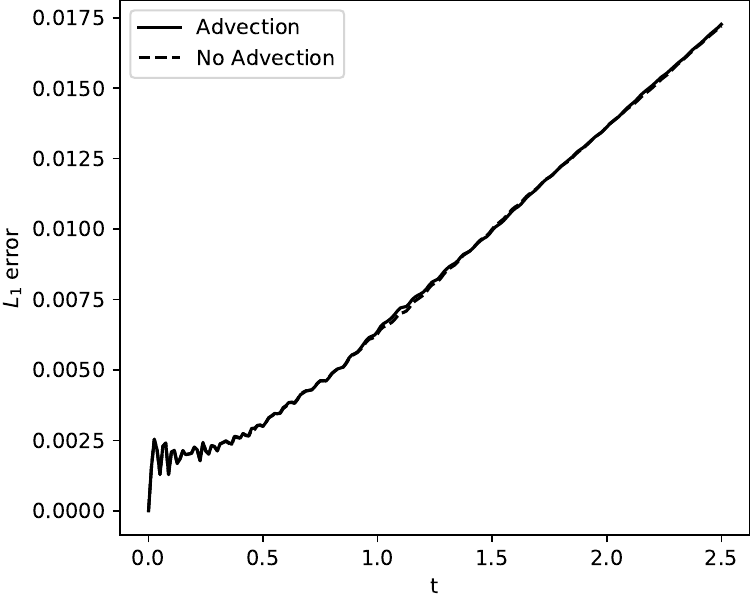}
    \subcaption{$L_1$ error in the velocity magnitude with time.}\label{fig:tg:advect:l1}
  \end{subfigure}
    \caption{Velocity decay plot for $Re = 100$ for both advection and no
      advection of particles in pseudo time as compared with the exact
      solution for $Re = 100$.}\label{fig:tg:advect}
\end{figure}
\begin{figure}[!h]
  \centering
  \begin{subfigure}[!h]{0.48\textwidth}
    \includegraphics[width=\linewidth]{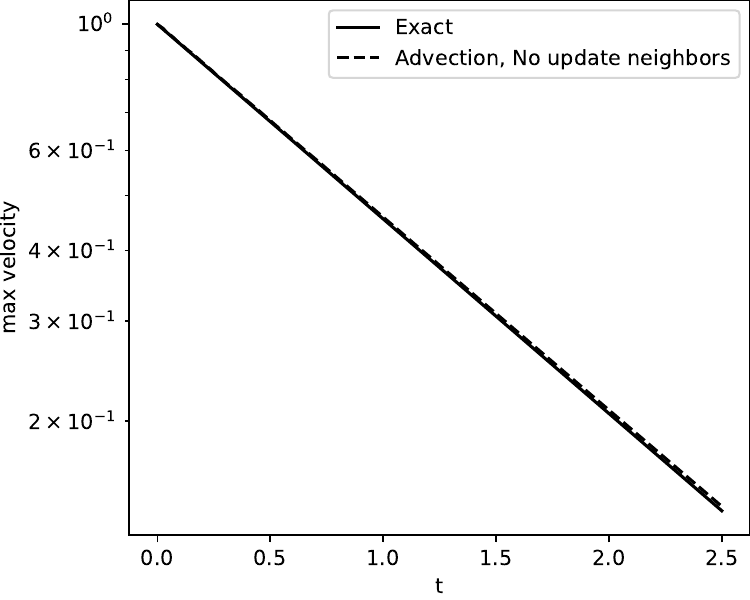}
    \subcaption{Decay of maximum velocity with time.}\label{fig:tg:advect:no_nnps:decay}
  \end{subfigure}
  \begin{subfigure}[!h]{0.48\textwidth}
    \centering
    \includegraphics[width=\linewidth]{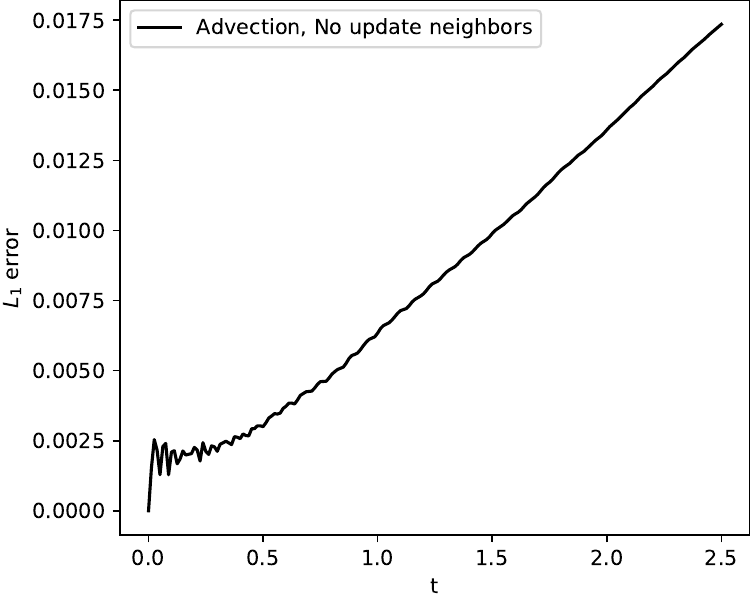}
    \subcaption{$L_1$ error in the velocity magnitude with time.}\label{fig:tg:advect:no_nnps:l1}
  \end{subfigure}
  \caption{Comparison for decay rates with time and $L_1$ errors in velocity
for advection and without advection cases while no update in the neighbour
particles.}
\label{fig:tg:advect:no_nnps}
\end{figure}
\begin{figure}[!h]
  \centering
  \includegraphics[width=\linewidth]{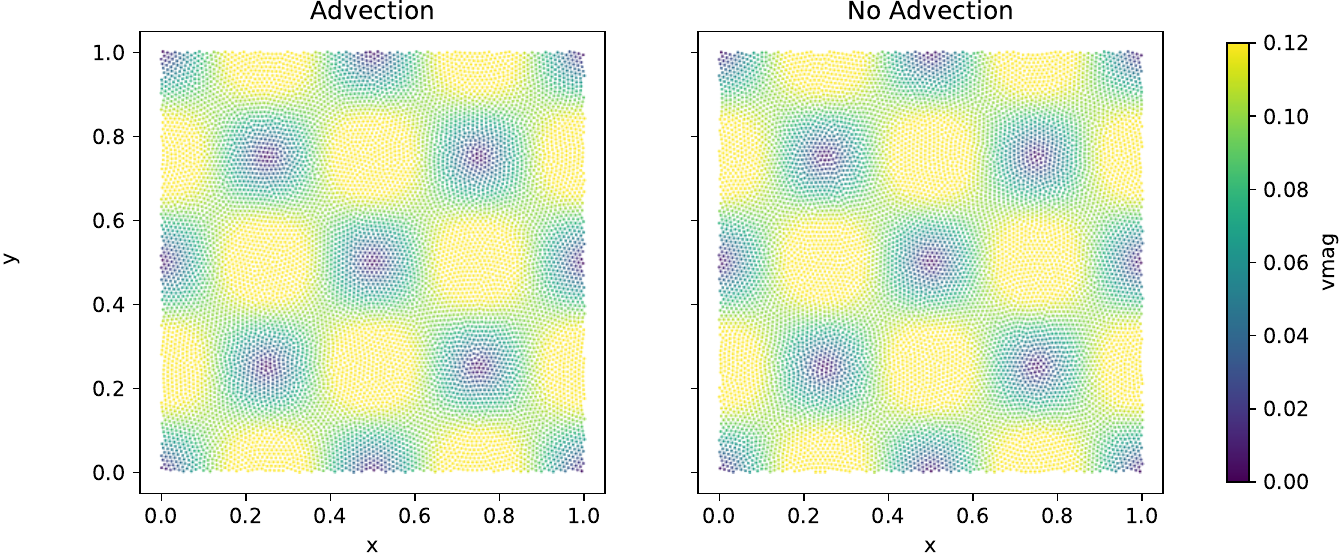}
  \caption{Particle plots for the Taylor-Green problem with advection and
    without advection in pseudo time while updating the
    neighbours.  Plots are shown at $t=2.5s$ with $Re=100$.}
  \label{fig:tg:advect:particle_plots}
\end{figure}

We recall that when we advect the particles in pseudo-time, we need to update
the neighbors, however the displacements are very small and this is not
necessary. We perform simulations to see if the differences are significant.
Fig.~\ref{fig:tg:advect:decay} shows the decay rate for the case with and
without advection in pseudo-time while updating the neighbours. There are no
noticeable differences in the results and the plots for each case lie on each
other. This is also seen in Fig.~\ref{fig:tg:advect:l1} which shows the $L_1$
error. Fig.~\ref{fig:tg:advect:no_nnps:decay} and
Fig.~\ref{fig:tg:advect:no_nnps:l1} show the decay rate and the $L_1$ error in
the velocity magnitude while not updating the neighbours resulting in a
similar conclusion that movement of particles in pseudo-time is too small to
significantly influence the results.

\begin{table}[!h]
\centering
\begin{tabular}{lr}
\toprule
                           Scheme &  CPU time (secs) \\
\midrule
                     DTSPH frozen &           326.87 \\
DTSPH advect, no update neighbors &           638.62 \\
   DTSPH advect, update neighbors &          1934.21 \\
\bottomrule
\end{tabular}

\caption{CPU time taken for $2.5$ secs of Taylor-Green simulation with
  $100 \times 100$ particles, with advection and without advection in pseudo
  time. }
\label{table:tg:advect:times}
\end{table}
While the accuracy is unaffected, the performance is significantly different
as can be seen from Table~\ref{table:tg:advect:times}. This shows that
advection of particles reduces performance by close to a factor of two. This
increase in performance is largely due to the fact that we re-calculate the
neighbour particles when we advect them. There is also some increase due to
the additional computations required for the advection.
Fig.~\ref{fig:tg:advect:particle_plots} shows the particle plots with color
representing velocity magnitude for the case where the particles are advected
and frozen. The results look identical. Based on these results, we do not
advect the particles in pseudo-time for any of the other simulations.

\subsubsection{The influence of $\beta$}
\label{sub:sec:tg:dtaufac}

The parameter $\beta$ is the ratio of $c/V_{ref}$, as discussed earlier. The
pseudo-timestep is also determined such that $\Delta t = \beta \Delta \tau$.
In this section we consider the Taylor-Green problem simulated at $Re=100$
using $100 \times 100$ particles, using the new scheme with different values of
$\beta$ chosen between 2 and 20 for a tolerance $\epsilon = 10^{-4}$, and run
for a simulation time of $t = 1$ sec.

\begin{figure}[!h]
  \centering
  \begin{subfigure}[b]{0.48\linewidth}
    \includegraphics[width=1.0\linewidth]{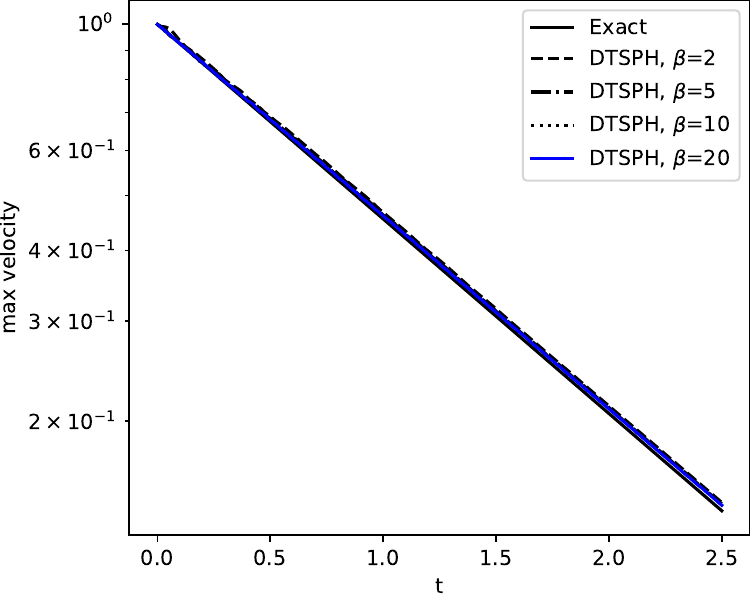}
    \subcaption{Decay of maximum velocity with time.}\label{fig:tg:beta:decay}
  \end{subfigure}
  \begin{subfigure}[b]{0.48\linewidth}
    \includegraphics[width=1.0\linewidth]{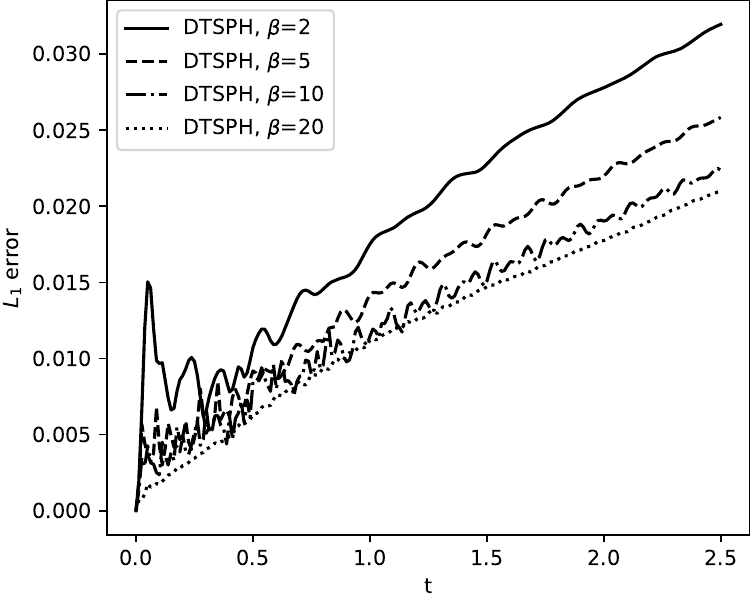}
    \subcaption{$L_1$ error of velocity magnitude with time.}\label{fig:tg:beta:l1}
  \end{subfigure}
  \caption{Decay rate of the maximum velocity and the $L_1$ error in the
    velocity magnitude for $\beta$ values of $[2, 5, 10, 20]$ for the DTSPH
    scheme with an error tolerance of $10^{-4}$.}\label{fig:tg:beta}
\end{figure}

Figs.~\ref{fig:tg:beta:decay} and~\ref{fig:tg:beta:l1} show the decay rate and
the $L_1$ error in the velocity for the different cases. From these it appears
that $\beta$ between 5 -- 20 works well.  We use a value of $\beta=10$ in all
simulations unless explicitly mentioned.
%
\subsubsection{Changing the convergence tolerance parameter $\epsilon$}
\label{sub:sec:tg:tol}

\begin{figure}[!h]
  \centering
  \begin{subfigure}[b]{0.48\linewidth}
    \includegraphics[width=1.0\linewidth]{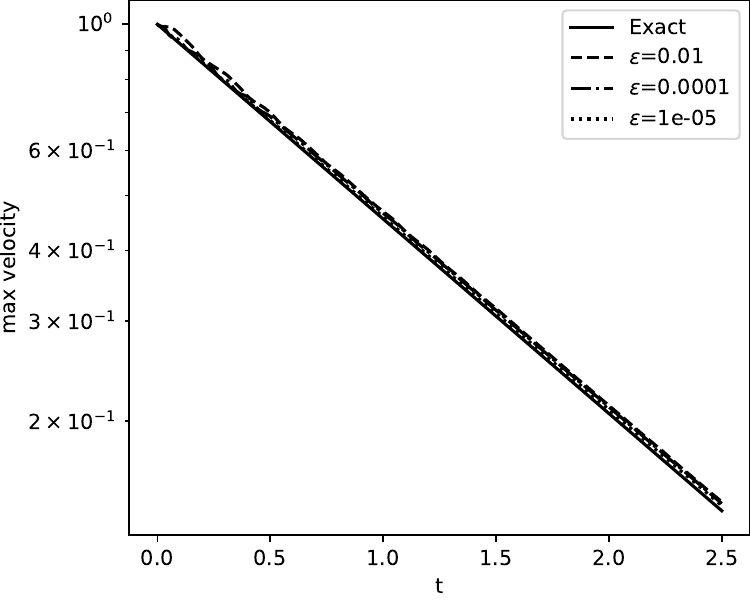}
  \subcaption{Decay of maximum velocity with time.}\label{fig:tg:tol:decay}
  \end{subfigure}
  \begin{subfigure}[b]{0.48\linewidth}
    \includegraphics[width=1.0\linewidth]{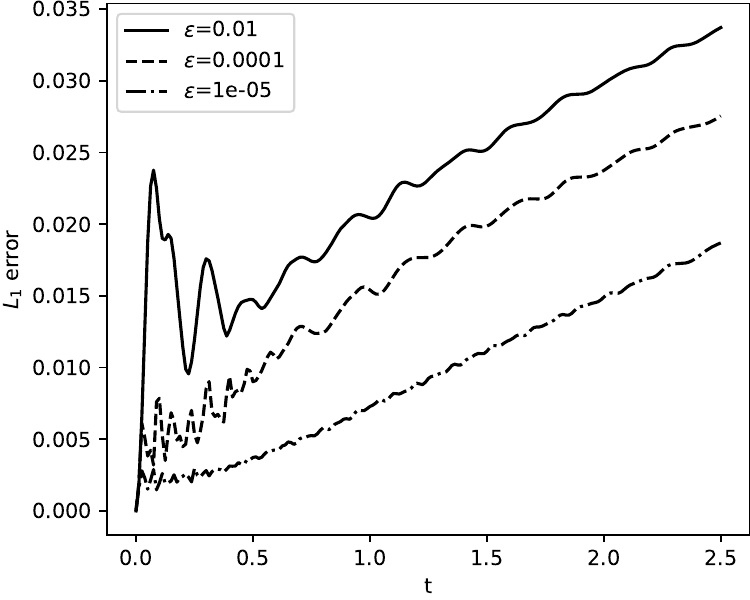}
  \subcaption{$L_1$ error in the velocity magnitude with time.}\label{fig:tg:tol:l1}
\end{subfigure}
\caption{Comparision for various tolerance $(\epsilon)$ ranging from
  $10^{-2}$ to $10^{-5}$.}
\end{figure}

We next choose $\beta=10$ and vary the tolerance from $10^{-2}$ to $10^{-5}$.
Fig.~\ref{fig:tg:tol:decay} shows the decay rates as the tolerance is changed
and Fig.~\ref{fig:tg:tol:l1} shows the $L_1$ error in the velocity. These
results clearly show that reducing the tolerance improves the accuracy.
However, we do see that the solutions are by-and-large robust to changes in
$\epsilon$ over a very large range. As expected, increase in the tolerance
leads to increase in the simulation time taken as seen in
Table.~\ref{table:tg:tol:times}.
\begin{table}[!htb]
\centering
\begin{tabular}{rr}
\toprule
$\epsilon$ &  CPU time (secs) \\
\midrule
      0.01 &           120.22 \\
    0.0001 &           133.87 \\
     1e-05 &           305.88 \\
\bottomrule
\end{tabular}

\caption{CPU time time taken for a simulation time of $2.5$ secs with $100 \times 100$
  particles with varying tolerance.}\label{table:tg:tol:times}
\end{table}
%
\subsubsection{Varying Reynolds number}
\label{sub:sec:tg:re}

We simulate the problem at $Re=1000$ using the most appropriate parameters
based on the previous results. The DTSPH scheme is used with a quintic spline
kernel, maximum initial random particle displacement of $\Delta x/ 10$, with a
tolerance of $\epsilon = 10^{-3}$, and $\beta = 5$ for different initial
particle arrangement of $25 \times 25$ to $200 \times 200$ and simulated for
$2.5$ secs.

\begin{figure}[!h]
  \centering
  \begin{subfigure}[b]{\linewidth}
    \includegraphics[width=1.0\linewidth]{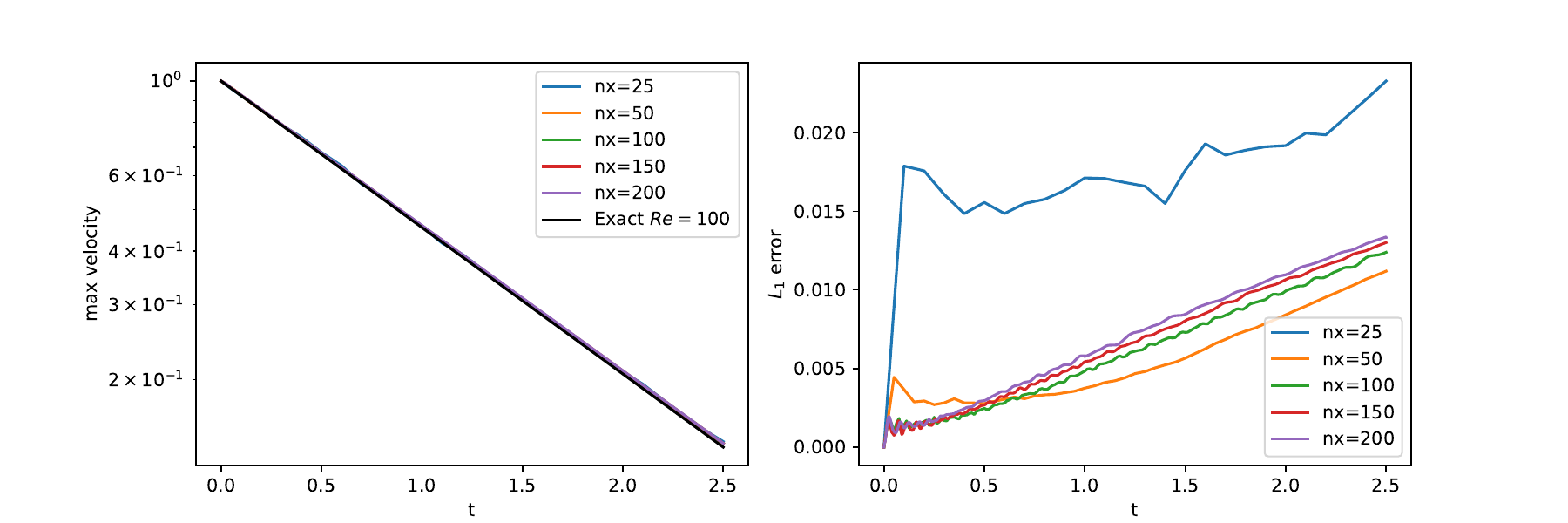}
  \subcaption{Decay of maximum velocity with time and the $L_1$ error in the
    velocity magnitude for $Re = 100$.}\label{fig:tg:re:100}
  \end{subfigure}

  \begin{subfigure}[b]{\linewidth}
    \includegraphics[width=1.0\linewidth]{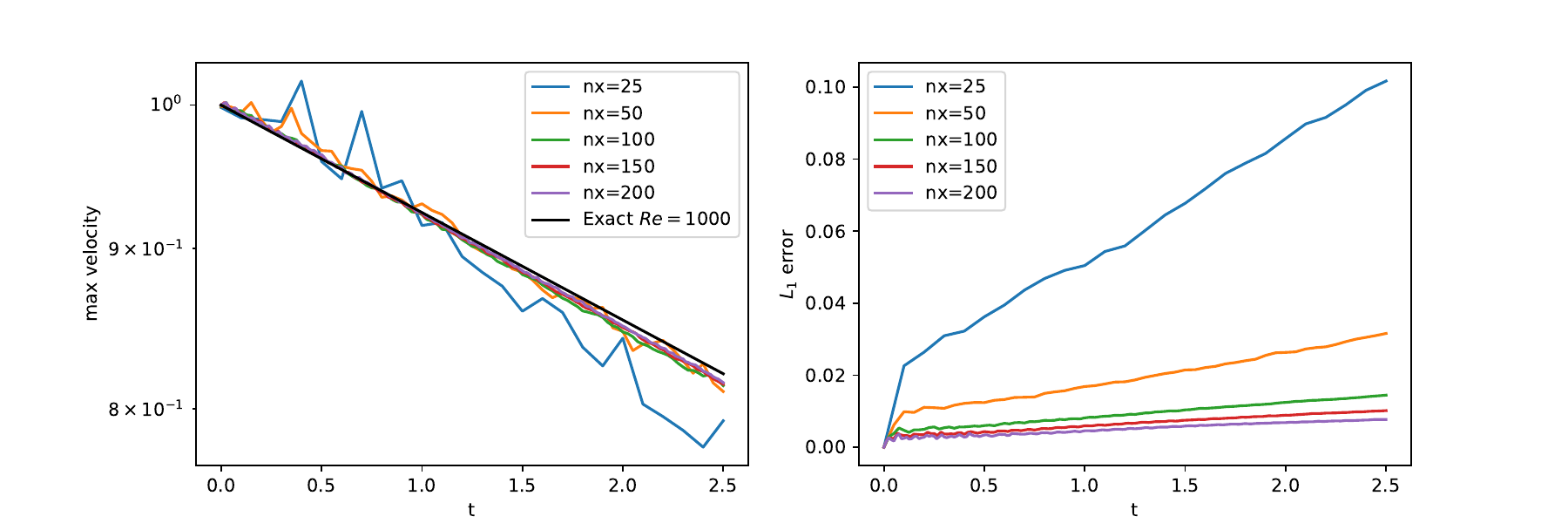}
  \subcaption{Decay of maximum velocity with time and the $L_1$ error in the
    velocity magnitude for $Re = 1000$.}\label{fig:tg:re:1k}
  \end{subfigure}
  \caption{Comparison of results for the Taylor-Green problem using five
    different particle configuration between $25 \times 25$ to $200 \times
    200$. Shown are the results for the Reynolds number of $Re = 100$ and
    $1000$.}\label{fig:tg:re}
\end{figure}

Fig.~\ref{fig:tg:re:100} shows the maximum velocity decay as well as the $L_1$
error of the velocity magnitude for the case of $Re=100$.
Fig.~\ref{fig:tg:re:1k} shows the same for $Re=1000$. It can be seen that in
the case of $Re=1000$ that there is a clear reduction in the errors as the
resolution is increased. In the case of $Re=100$, it appears that the errors
are lower for the $50 \times 50$ resolution and increase by a small amount as
the resolution is increased. We believe that this occurs because of possible
issues with the convergence of the discretization of the diffusive terms in
the governing equations. We point out that at $Re=500$ we obtain similar
convergence as in the case of $Re=1000$ showing that issue is not with the
convergence of the DTSPH scheme per-se. These results show that the new scheme
performs very well.

\subsubsection{Comparision with other schemes}
\label{sub:sec:tg:compare}

Here we simulate the problem for $t = 2.5s$ for $Re = 100$ using the new DTSPH
scheme comparing it with WCSPH, $\delta$-SPH, and EDAC. The quintic spline
kernel is used for all the schemes with $h = \Delta x$. For all the cases the
particles are perturbed by atmost $\Delta x / 10$. For DTSPH, we use
$\beta = 10$ with a tolerance of $\epsilon = 10^{-4}$. We use an initial
configuration of $100 \times 100$ particles.

\begin{figure}[!h]
  \centering
  \begin{subfigure}[b]{0.48\linewidth}
    \includegraphics[width=1.0\linewidth]{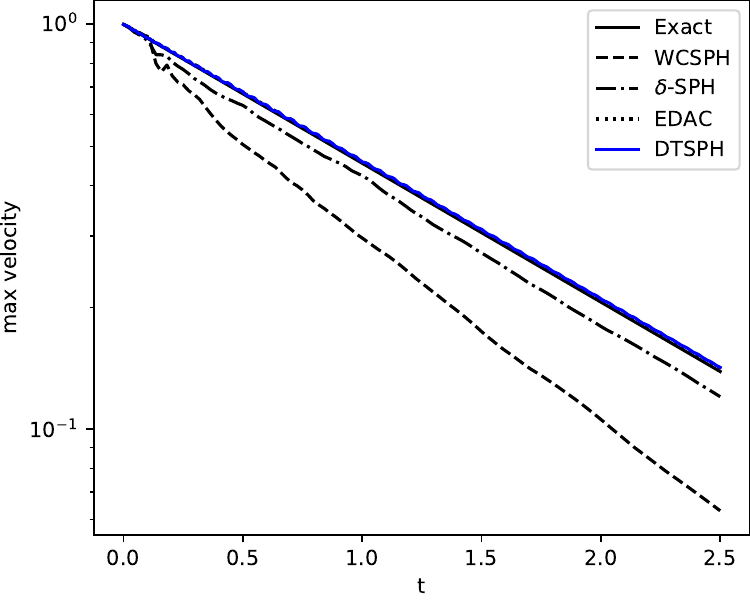}
  \subcaption{Decay of maximum velocity with time.}\label{fig:tg:re:decay}
  \end{subfigure}
  \begin{subfigure}[b]{0.48\linewidth}
    \includegraphics[width=1.0\linewidth]{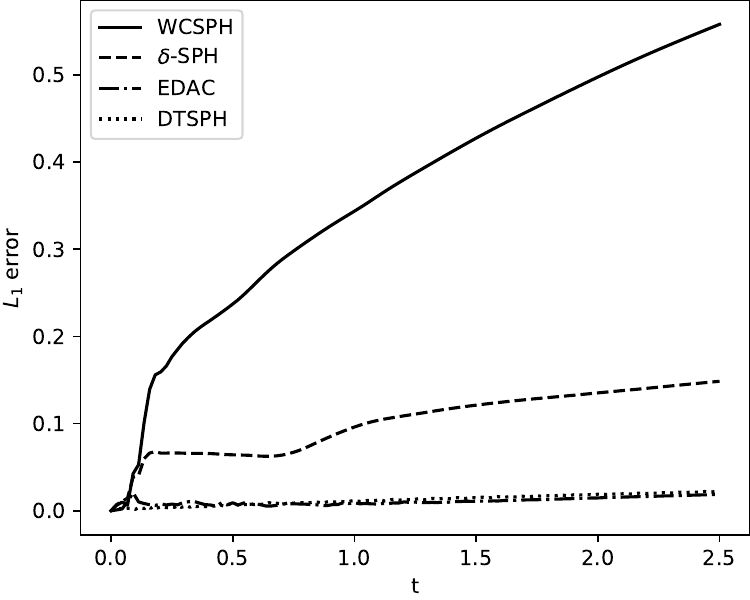}
  \subcaption{$L_1$ error in the velocity magnitude with time.}\label{fig:tg:re:l1}
  \end{subfigure}
  \caption{Comparision of DTSPH with other schemes for the simulation of
    Taylor-Green problem, with $Re = 100$ and using $100 \times 100$
    particles.}\label{fig:tg:compare}
\end{figure}

As can be seen from the results shown in Fig.~\ref{fig:tg:compare}, the new
scheme is more accurate than the standard WCSPH scheme. The scheme is more
accurate than the standard $\delta$-SPH scheme. We note that we do not employ
any form of shifting for the $\delta$-SPH and WCSPH scheme cases. The scheme
is not more accurate than the EDAC scheme as in the EDAC scheme, the particles
are also regularized using the transport velocity formulation which
significantly improves the results. As can be seen from the
Table~\ref{table:tg:compare:times}, the new scheme is anywhere from 1.9 to 2.8
times faster than the other schemes.

\begin{figure}[!h]
  \centering
  \includegraphics[width=\linewidth]{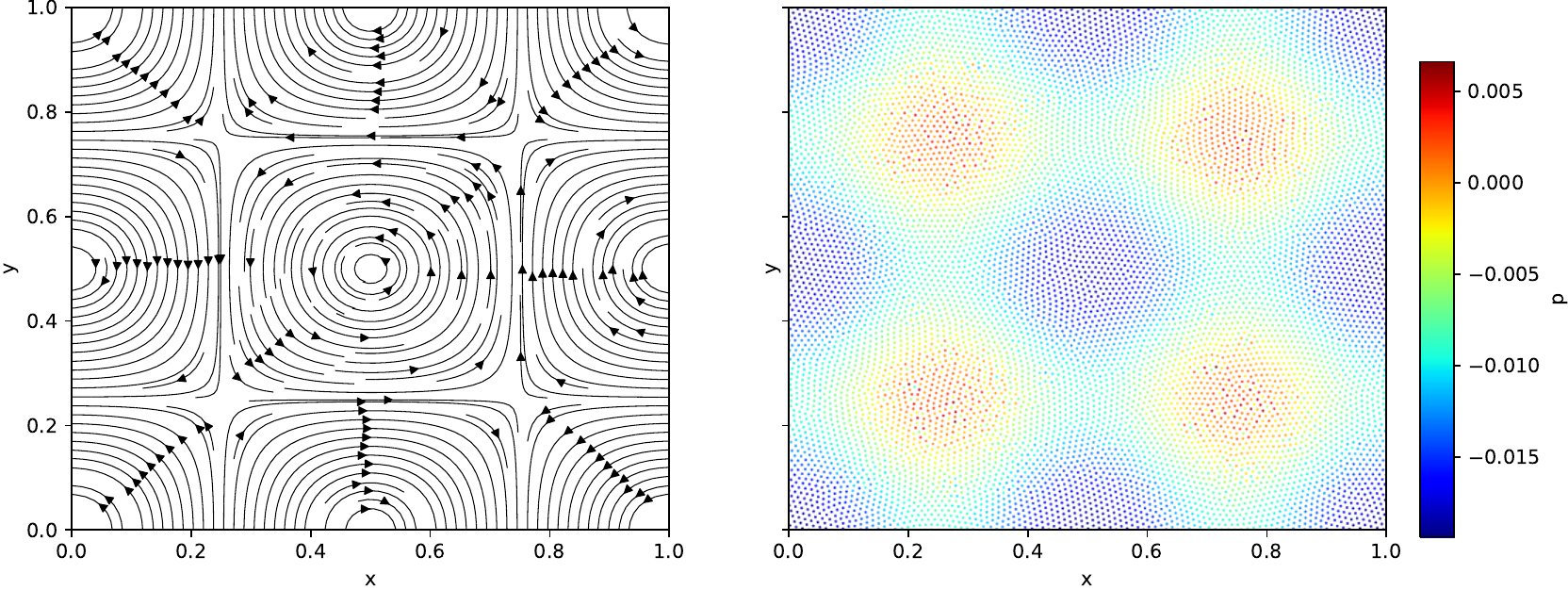}
  \caption{Particle plots for the Taylor-Green problem, with $100 \times 100$
    particles in the initial configuration, showing streamlines on the
    left and pressure on the left at $t = 2.5s$.}%
  \label{fig:tg:streamplot}
\end{figure}

In Fig.~\ref{fig:tg:streamplot}, we show the streamlines as well as the
particle plot with the color indicating pressure at $t=2.5$secs for the DTSPH
case for $Re=100$.  As can be seen the particle distribution is smooth.
\begin{table}[!htb]
\centering
\begin{tabular}{lr}
\toprule
      Scheme &  CPU time (secs) \\
\midrule
       DTSPH &           124.49 \\
       WCSPH &           239.70 \\
        EDAC &           275.43 \\
$\delta$-SPH &           347.98 \\
\bottomrule
\end{tabular}

\caption{CPU time time taken for a simulation time of $2.5$ secs with $100 \times 100$
  particles for various schemes.}\label{table:tg:compare:times}
\end{table}

\subsection{The lid-driven-cavity problem}
\label{sec:ldc}

\begin{figure}[!h]
  \centering
  \includegraphics[width=\linewidth]{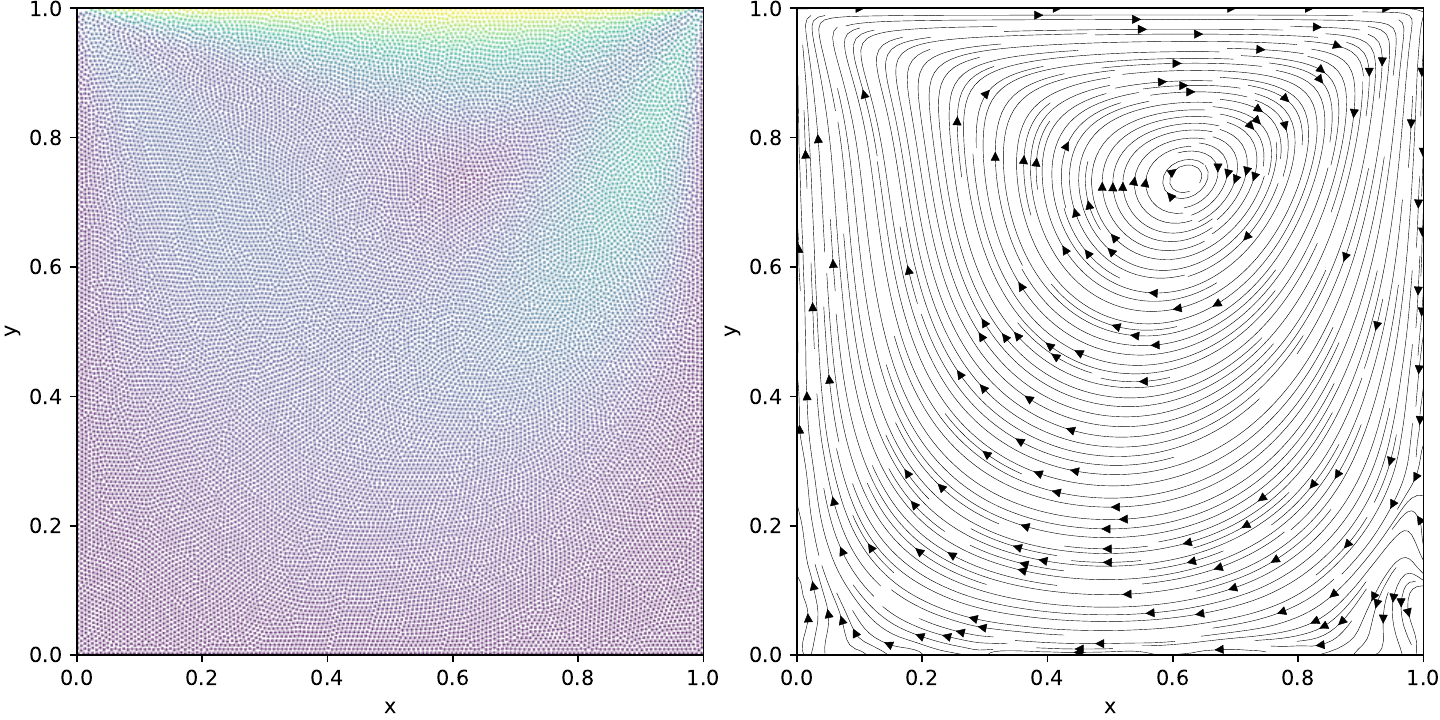}
  \caption{Particle plots for the cavity problem, with $150 \times 150$
    particles in the initial configuration, showing the velocity magnitude on
    the left and streamlines on the right at $t = 10s$.}%
  \label{fig:ldc:streamplot}
\end{figure}

We next consider the classic lid-driven-cavity problem. This is a fairly
challenging problem to simulate with SPH~\cite{Adami2013, ghasemi-2013,
  lee-comparison-2008}. The fluid is placed in a unit square with a lid moving
with a unit speed to the right. The bottom and side walls are treated as
no-slip walls. The Reynolds number of the problem is given by
$Re=\frac{V}{\nu}$, where $V$ is the lid velocity. We use a quintic spline
kernel with $h=\Delta x$. The problem is simulated at $Re=100$ using a
$50\times 50$, $100\times 100$, and $150 \times 150$ grid for a simulation
time of $t = 10s$ until there is no change in the kinetic energy of the
system. For the DTSPH scheme we use a $\beta = 10$ with a tolerance
$\epsilon=10^{-4}$. The results are compared with those of the TVF
scheme\cite{Adami2013} and the established results of Ghia et
al.~\cite{ldc:ghia-1982}.
\begin{figure}[!h]
  \centering
  \includegraphics[width=\linewidth]{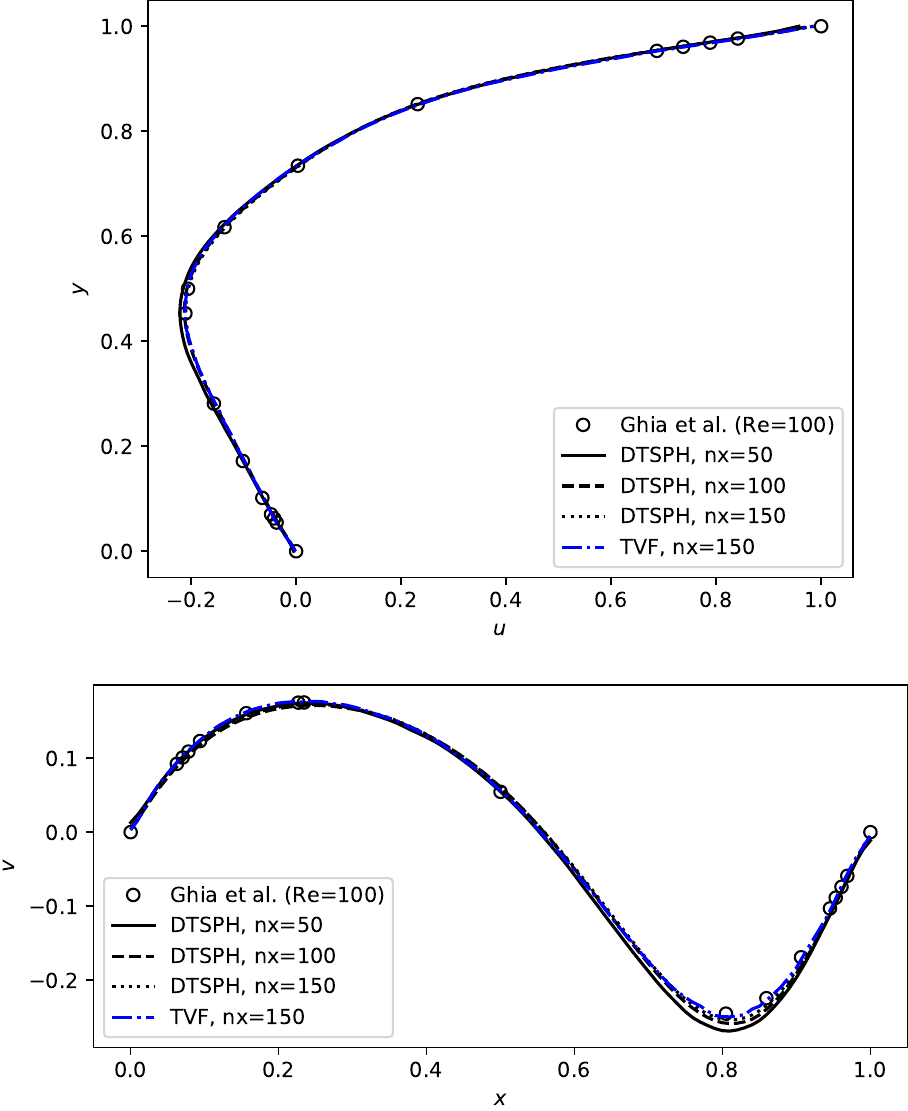}
  \caption{Velocity profiles $u$ vs.\ $y$ and $v$ vs.\ $x$ for the
    lid-driven-cavity problem at $Re=100$ with three initial particle
    arrangement of $50 \times 50$, $100 \times 100$, and $150 \times
    150$. Here we compare DTSPH with TVF and the results
    of~\cite{ldc:ghia-1982}.}\label{fig:ldc:uv_re100}
\end{figure}
Fig.~\ref{fig:ldc:uv_re100}, shows the centerline velocity profiles for $u$
vs.\ $y$ and $v$ vs.\ $x$ for different resolutions of particles. It is seen
that the TVF scheme produces better results as expected. However, the results
of the new scheme are in good agreement. In Fig.~\ref{fig:ldc:streamplot}, we
show the particle distribution with the velocity magnitude on the left and
streamlines on the right.

\subsubsection{Steady Lid-driven cavity}
\label{sec:steady_ldc}

In order to show that we are able to obtain steady state results, we employ
the steady state equations discussed in Section~\ref{sec:steady-state} to
solve the lid-driven-cavity problem. We solve the problem until there is no
change in the kinetic energy of the system. We simulate the problem using a
quintic spline kernel for $Re=100$ and $Re=1000$ using a $50\times 50$,
$100\times 100$ and $150 \times 150$ particle grid. For $Re=100$ we simulate
the problem up to $\tau=10$ and for $Re=1000$ we simulate up to $\tau=50$.
\begin{figure}[!h]
  \centering
  \includegraphics[scale=0.5]{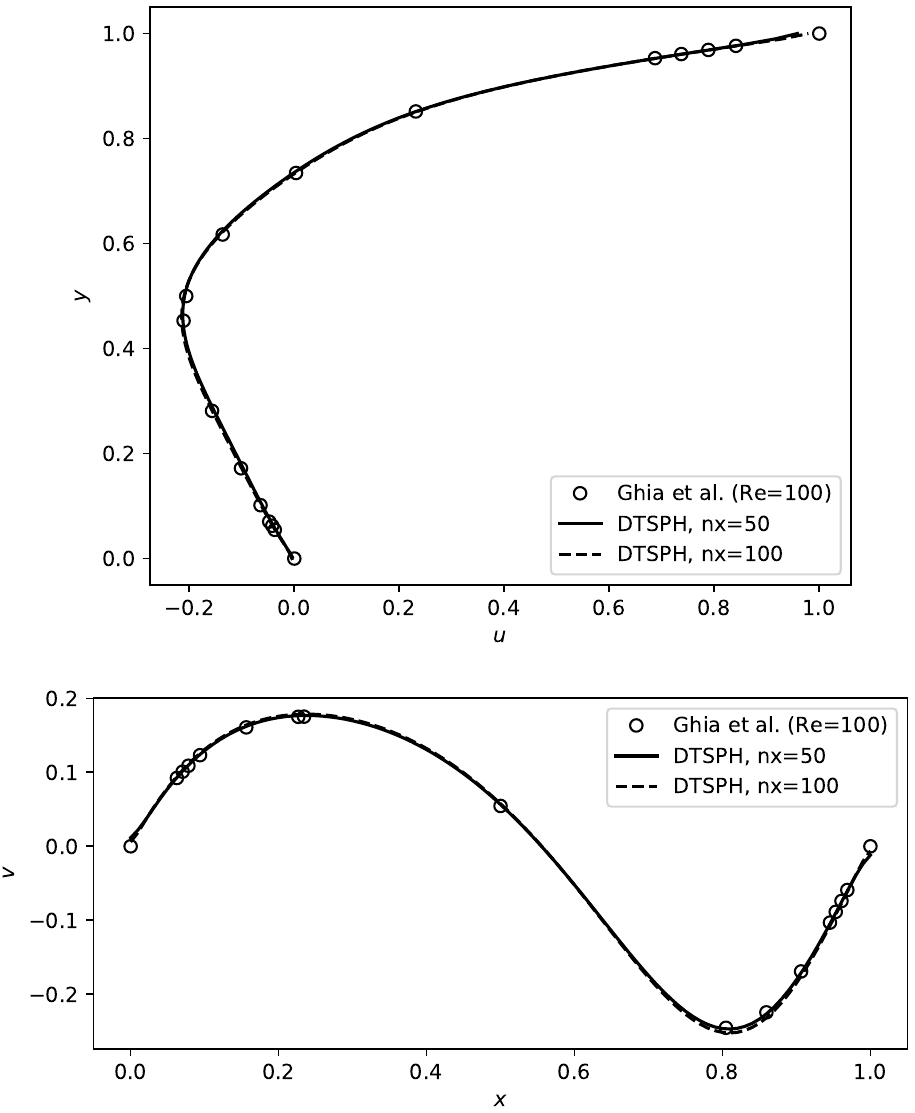}
  \caption{Velocity profiles for the lid-driven-cavity using the steady state
    simulation procedure for $Re = 100$ with initial partial arragement of
    $50 \times 50$, $100 \times 100$, and $150 \times 150$ compared with the
    results of~\cite{ldc:ghia-1982}.}
\label{fig:steady:ldc:uv_re100}
\end{figure}
\begin{figure}[!h]
  \centering
  \includegraphics[width=0.4\linewidth]{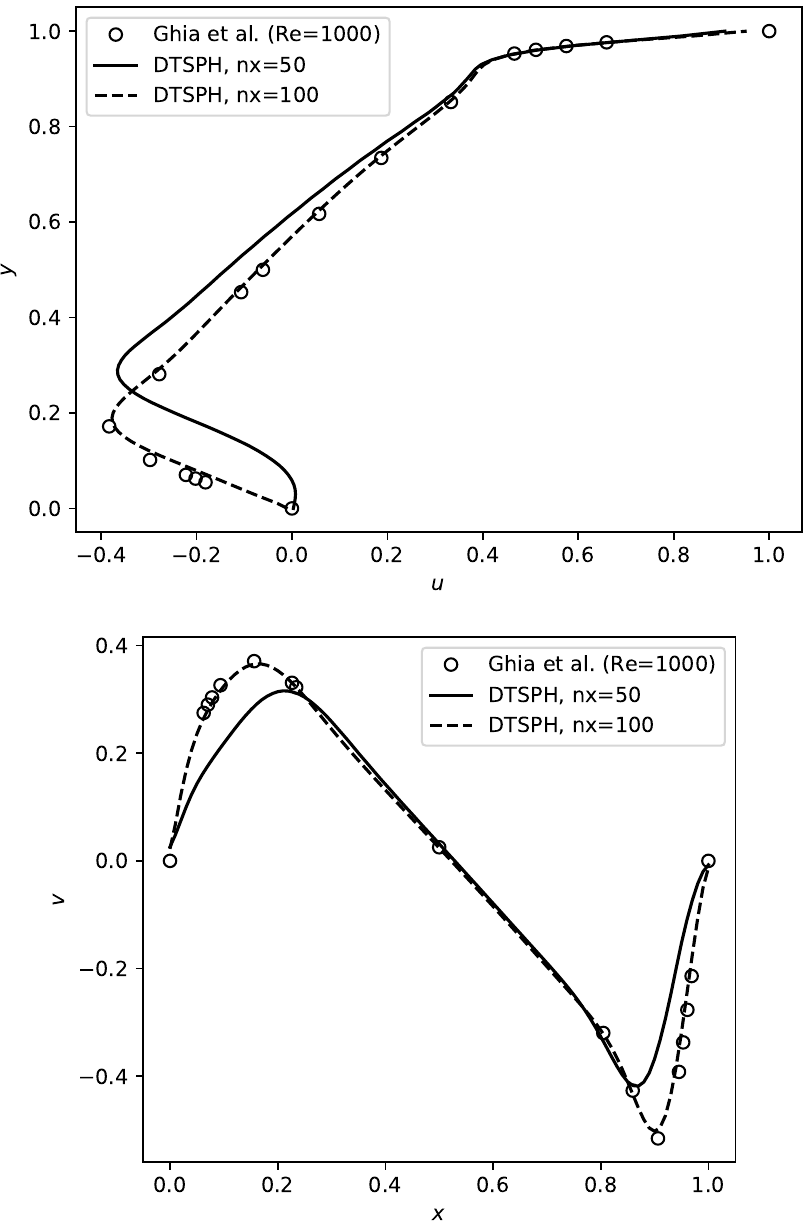}
  \caption{Velocity profiles for the lid-driven-cavity using the steady state
    simulation procedure for $Re = 1000$ with initial partial arragement of
    $50 \times 50$, $100 \times 100$, and $150 \times 150$ compared with
    the results of~\cite{ldc:ghia-1982}.}
\label{fig:steady:ldc:uv_re1000}
\end{figure}
Fig.~\ref{fig:steady:ldc:uv_re100} shows the velocity profiles for the
$Re=100$ case and Fig.~\ref{fig:steady:ldc:uv_re1000} shows velocity profiles
for the $Re = 1000$ case.

These results show that we are able to simulate internal flows very well using
the new DTSPH scheme. We have also demonstrated that the steady-state
equations also work very well. We next consider problems that involve a
free-surface.
\subsection{Square patch}
\label{sec:sp}

The square patch problem~\cite{colagrossi-phdthesis:2005, khayyer-2013,
  sun-deltap-2017} is a free surface problem where a square patch of fluid of
side $L$ is subjected to the following initial conditions,
\begin{equation}
  \begin{aligned}
    u_{0}(x, y) &= \omega y, \\
    v_{0}(x, y) &= -\omega x,
  \end{aligned}
\end{equation}
\begin{equation}
  p_{0}(x, y) = \rho \sum_m^\infty \sum_n^\infty -\frac{32\omega^2/(m n \pi^2)}
  {\left[{\left(\frac{n\pi}{L}\right)}^2 + {\left(\frac{m \pi^2}{L}\right)}^2
    \right]} \sin\left(\frac{m \pi x^*} {L}\right)\sin\left(\frac{n \pi y^*}
    {L}\right) m, n \in \mathbb{N}_{odd},
\end{equation}
where $X^* = x + L/2$ and $y^* = y + L/2$.

We simulate this problem for $t = 3s$ using the DTSPH, and EDAC schemes for
comparison. In this case, the EDAC simulations also employ particle
shifting~\cite{diff_smoothing_sph:lind:jcp:2009}. The quintic spline kernel
with $h/\Delta x = 1.3$ is used for all the schemes, artificial viscosity
$\alpha = 0.1$ is used for all the schemes. For the DTSPH scheme, $\beta =
10$~with a tolerance $\epsilon = 10^{-3}$ is used. Two different initial
configurations of $50 \times 50$ and $200 \times 200$ particles are used.
\begin{figure}[!h]
  \centering
  \includegraphics[width=1.0\textwidth]{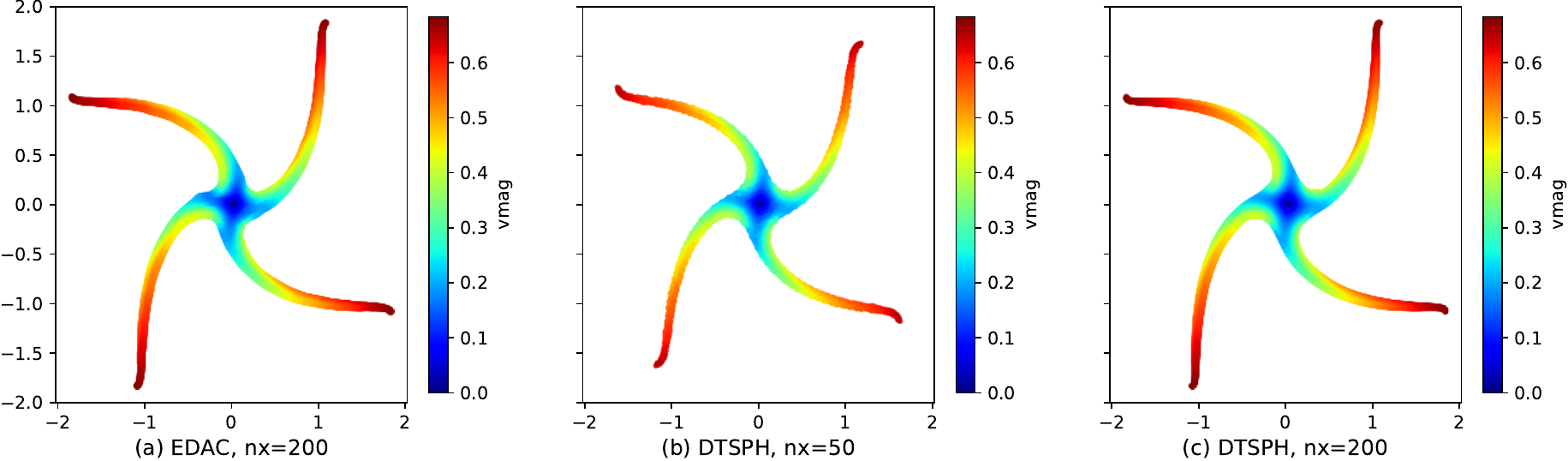}
  \caption{Particle distribuiton plots at $t = 3$~secs for the square patch
    problem. Artificial viscosity is used in all the schemes. Top row
    corresponds to $50 \times 50$ particles, and the bottom row corresponds to
    $100 \times 100$ particles. In column (a) EDAC scheme is used, with
    $200 \times 200$ particles, column (b) indicates DTSPH scheme with a
    tolerance of $\epsilon = 10^{-3}$, and $50 \times 50$ particles, and in
    column (c) DTSPH scheme is used with a tolerance of $\epsilon = 10^{-3}$,
    and $200 \times 200$ particles.}
\label{fig:sp:particle_plots}
\end{figure}
The particle distribution for each scheme at the end of $t=3s$ is shown in
Fig.~\ref{fig:sp:particle_plots}. The plots of DTSPH and EDAC are in good
agreement with each other showing that the new scheme is as good as the EDAC
scheme.
\subsection{Elliptical drop}
\label{sec:ed}

The elliptical drop problem was first solved in the context of the SPH by
Monaghan~\cite{sph:fsf:monaghan-jcp94}. This problem is also solved in the
context of truly incompressible SPH~\cite{khayyer-2013, lind-2016,
  rezavand-2018a}. In this problem an initially circular drop of inviscid
fluid having unit radius is subjected to the initial velocity field given by
$-100x\hat{i} + 100 y \hat{j}$. The outer surface is treated as a free
surface. Due to the incompressibility constraint on the fluid there is an
evolution equation for the semi-major axis of the ellipse.

This problem is simulated using the DTSPH, $\delta$-SPH (without shifting),
and EDAC (with shifting) respectively. An artificial viscosity parameter of
$\alpha = 0.15$ is used for all the schemes. An error tolerance of $\epsilon =
10^{-4}$ is used for the DTSPH and scheme. $\beta=10$, $\Delta x=0.02$, $h=1.3
\Delta x$ and a quintic spline kernel is used for all the schemes. The
simulation is run for $t = 0.0076s$.

\begin{figure}[!h]
  \centering
  \includegraphics[width=0.8\textwidth]{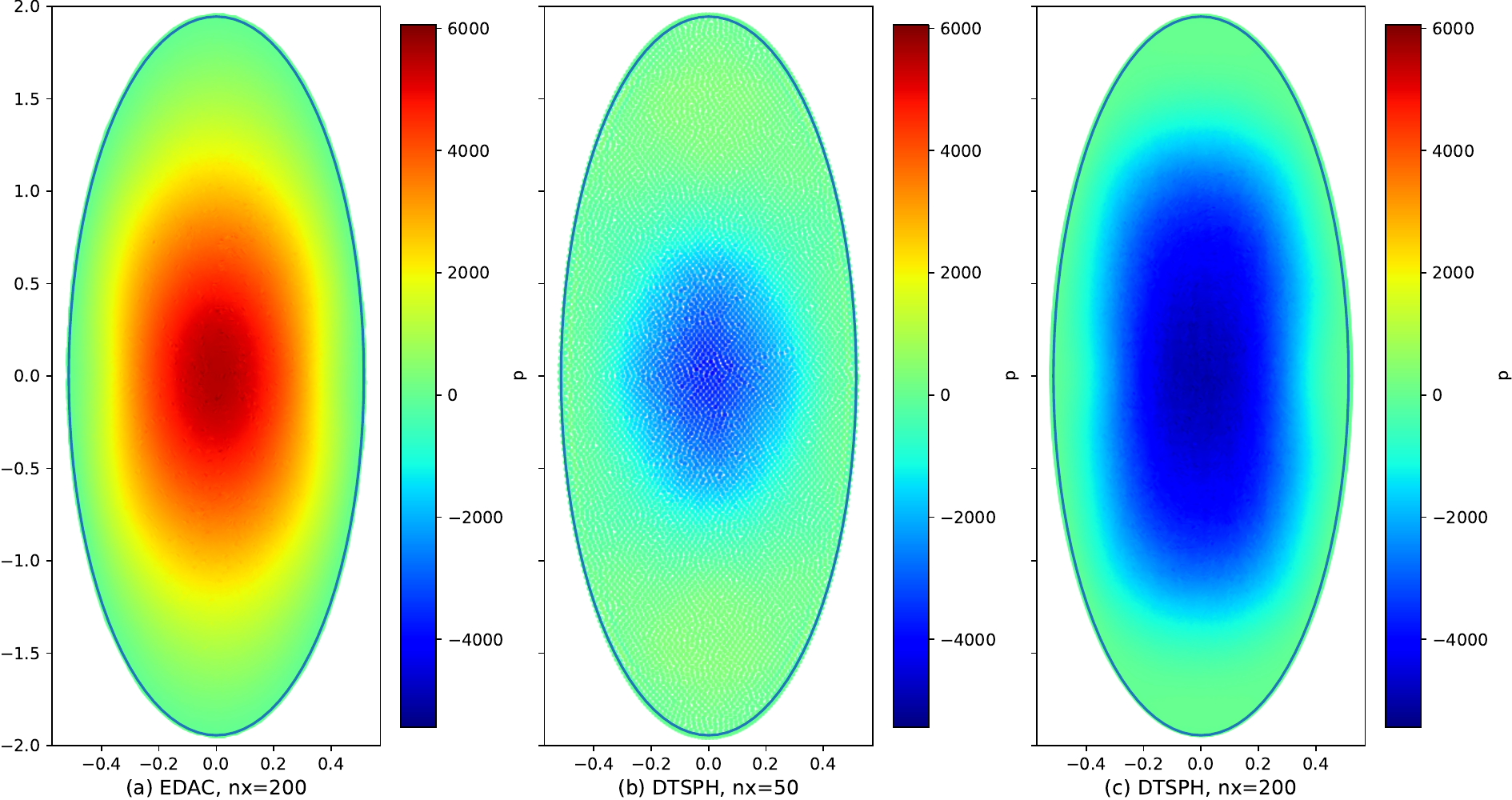}
  \caption{The distribution of particles for the elliptical drop problem at $t
    = 0.0076$ seconds. The plot (a) is with the EDAC using $200\times 200$
    particles. Plot (b) is that of the new DTSPH scheme with $50 \times 50$
    particles, (c) uses DTSPH with $200 \times 200$. The solid blue line is
    the exact solution for the shape of the drop and the colors indicate the
    pressure.}
\label{fig:ed:particle_plots}
\end{figure}

Fig.~\ref{fig:ed:particle_plots} shows the distribution of particles for
different schemes. The colors indicate the pressure. As can be seen, the
DTSPH and EDAC results are similar. It is important to note that all
the pressure values are in a similar range with none of the schemes exhibiting
severe noise in the pressure.
\begin{figure}[!h]
  \centering
  \includegraphics[width=0.6\linewidth]{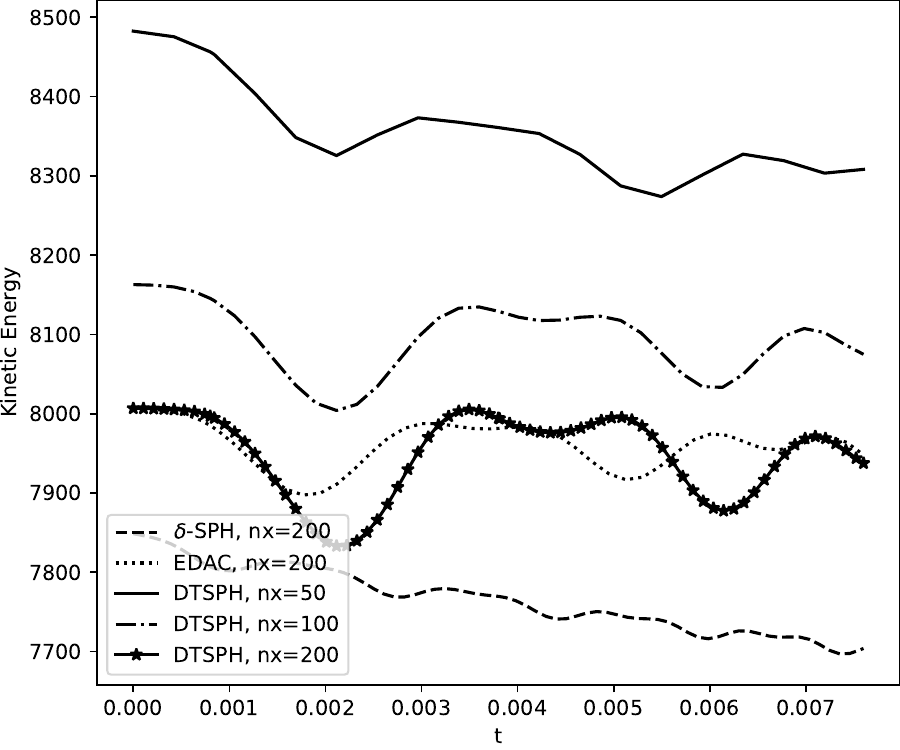}
  \caption{The kinetic energy with time of the Elliptical drop problem as
computed with DTSPH, $\delta$-SPH, and EDAC schemes.}
\label{fig:ed:ke}
\end{figure}
\begin{figure}[!h]
  \centering
  \includegraphics[width=0.6\linewidth]{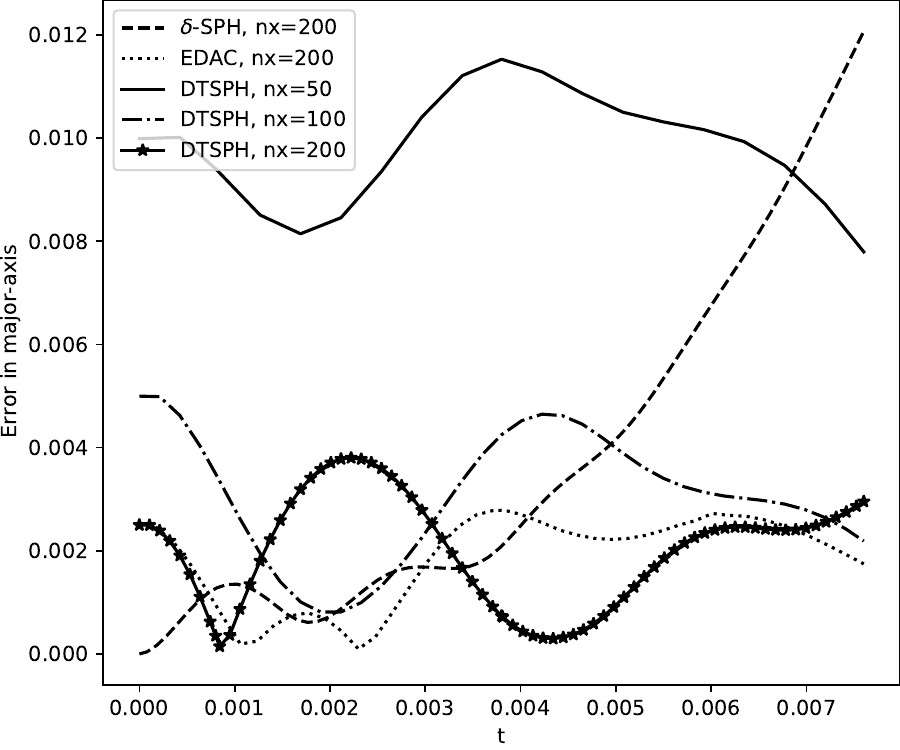}
  \caption{Error in computed size of semi-major axis of the elliptical drop
    problem compared with the exact solution for the DTSPH, $\delta$-SPH and EDAC schemes.}
\label{fig:ed:major_axis}
\end{figure}
Fig.~\ref{fig:ed:ke}, shows the evolution of the kinetic energy, the results
are similar for DTSPH and EDAC schemes. This is to be expected. However it is
interesting to note that the kinetic energy of the $\delta$-SPH scheme decays
faster than the other schemes. Fig.~\ref{fig:ed:major_axis} shows the error in
the semi-major axis as compared to the exact solution. The DTSPH and EDAC both
perform slightly better than the $\delta$-SPH scheme. The results indicate
that the new scheme performs well in comparison with state of the art
weakly-compressible schemes.

\subsection{Dam-break in 2 dimensions}
\label{sec:db2d}

A two dimensional dam-break over a dry bed~\cite{lee-comparison-2008,
  marrone-deltasph:cmame:2011, lind-2016} is considered next. The DTSPH and the
standard EDAC schemes are compared. The simulation is performed for $1s$.  The
quintic spline kernel is used with $h/\Delta x = 1.0$, and an artificial
viscosity of $\alpha = 0.1$ is used for all the schemes. A tolerance of
$\epsilon=10^{-4}$ is used for DTSPH.

The problem considered is described in~\cite{lee_violeau:db3d:jhr2010} with a
block of fluid column of height $h = 2m$, width $w = 1m$. The block is
released under gravity which is assumed to be $-9.81 m/s^2$.
\begin{figure}[!h]
  \centering
  \includegraphics[width=1.0\linewidth]{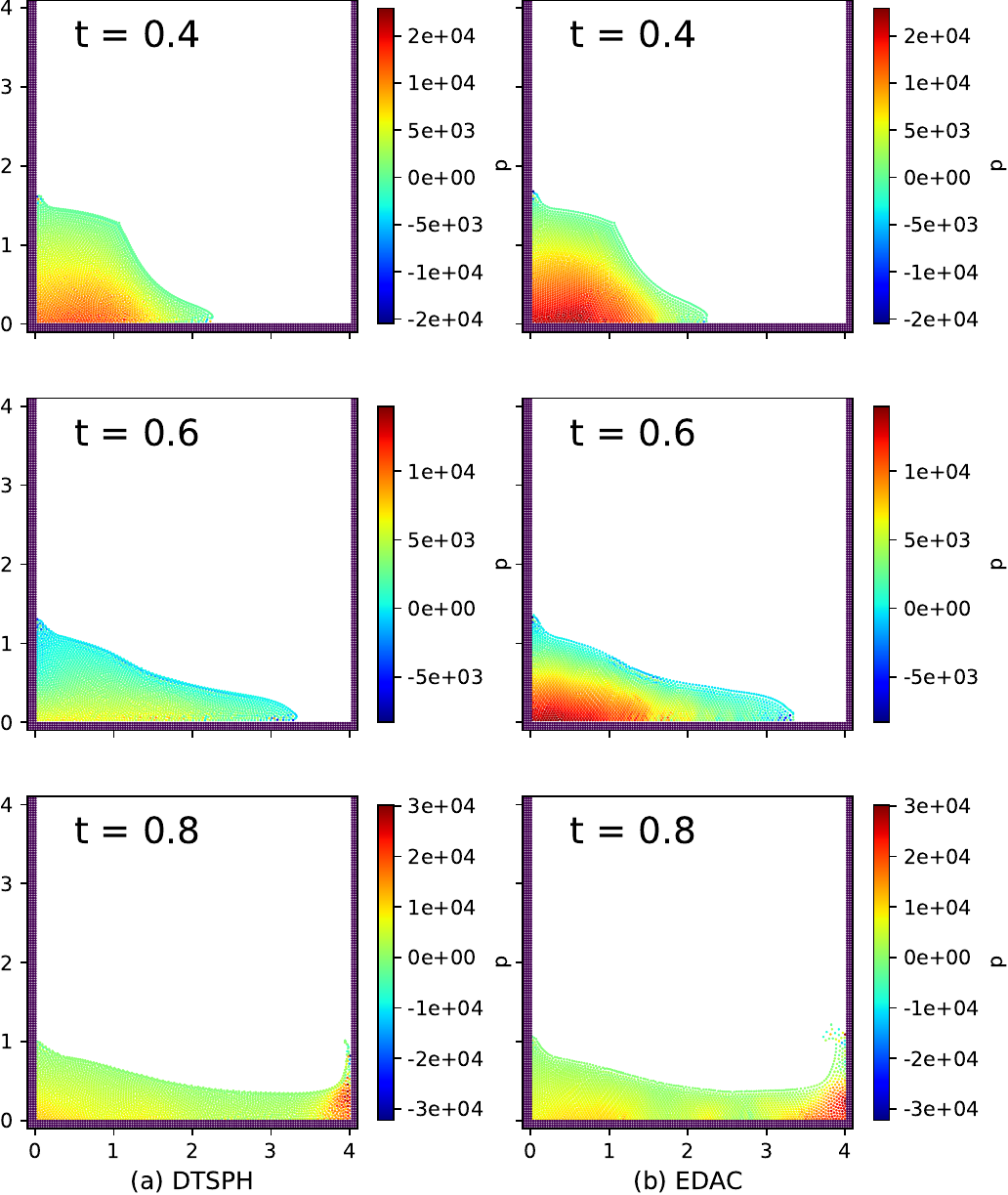}
  \caption{Particle distribution plots with color indicating pressure for the
    dam-break 2D problem at various times. DTSPH is shown on the left, and
    EDAC is shown on the right. Top row is at $t = 0.4$ secs, second row is at
    $t = 0.6$ secs, and bottom row is at $t = 0.8$ secs.}
\label{fig:db:particle_plots:pre}
\end{figure}

\begin{figure}[!h]
  \centering
  \includegraphics[width=1.0\linewidth]{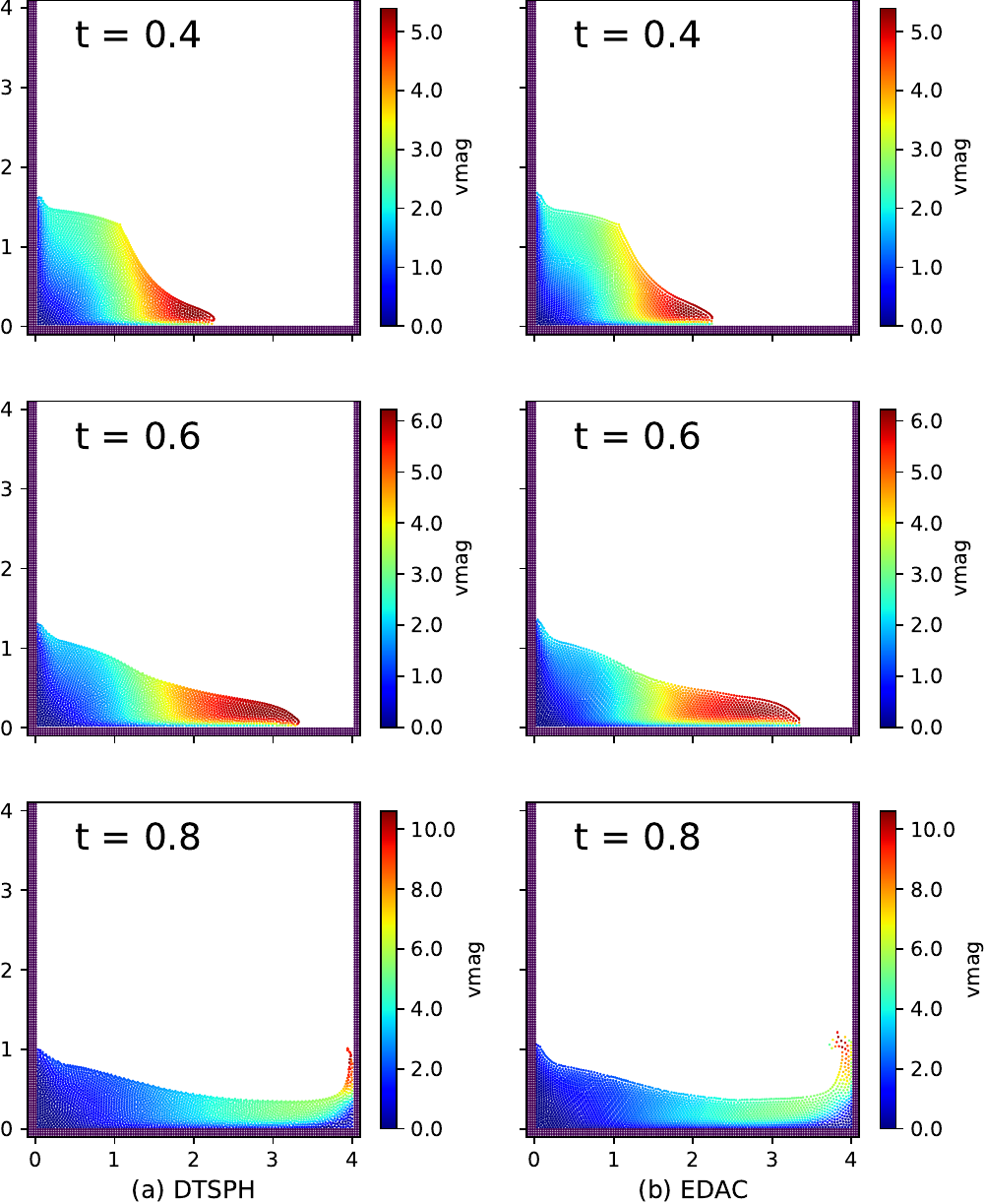}
  \caption{Particle distribution plots with color indicating velocity
    magnitude for the 2D dam-break problem at variuos times. DTSPH scheme is
    shown on the left, and EDAC is shown on the right. Top row is at $t = 0.4$
    secs, the second row is at $t=0.2$ secs, and bottom row is at $t = 0.8$
    secs.}
\label{fig:db:particle_plots:vmag}
\end{figure}
\begin{figure}[!h]
  \centering
  \includegraphics[width=0.6\linewidth]{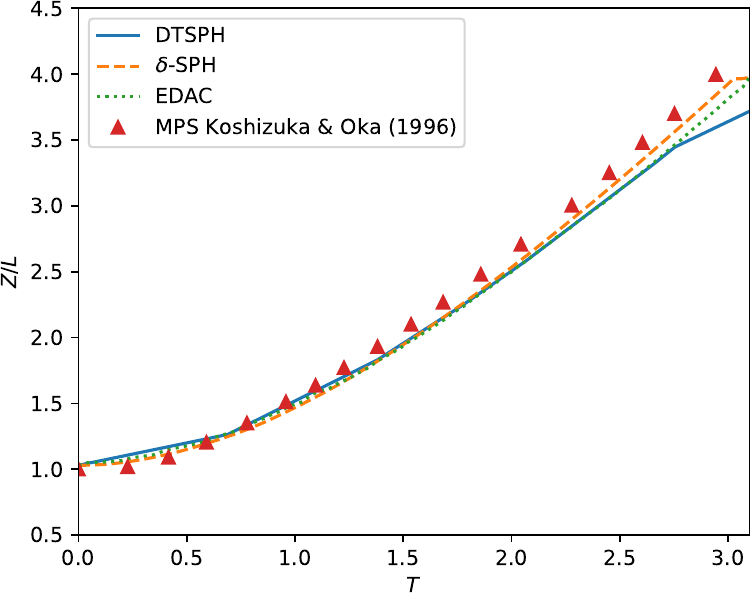}
  \caption{Position of the toe of the dam versus time of DTSPH, WCSPH,
    $\delta$-SPH and EDAC, as compared with the simulation
    of~\cite{koshizuka_oka_mps:nse:1996}. $Z$ is the distance of toe of the
    dam from the left wall and $L$ is the initial width of the dam}
\label{fig:sp:wcsph_toe_vs_t}
\end{figure}
For the DTSPH and EDAC schemes, the particle distribution is shown in
Fig.~\ref{fig:db:particle_plots:pre} at various times with color indicating
pressure. Fig.~\ref{fig:db:particle_plots:vmag} shows the particle
distribution with color indicating velocity magnitude at various times. The
results of the new scheme seem largely comparable with that of the EDAC
scheme. The results also show the improvements obtained by the addition of
diffusive term in the pressure evolution equation. This significantly reduces
the noise. The standard EDAC results are very similar to those of the
$\delta$-SPH and are hence not shown. Fig.~\ref{fig:sp:wcsph_toe_vs_t} plots
the position of the toe of the dam versus time as compared with the results of
the Moving Point Semi-implicit scheme of~\cite{koshizuka_oka_mps:nse:1996}.
The results of the DTSPH scheme are in good agreement with those of the EDAC
scheme.

\subsection{Dam-break in three dimensions}
\label{sec:db3d}

A three dimensional case is shown to demonstrate the performance of the new
scheme as compared to the EDAC scheme. This is an important case as the
previous problems only require a smaller number of particles. We consider a
three-dimensional dam break over a dry bed with an obstacle. A cubic spline
kernel is used for both the new scheme and the EDAC with $h/\Delta x = 1.3$
and artificial viscosity $\alpha=0.1$. The problem is simulated for a total
time of 1 second.  We do not use any particle shifting in this case.

The problem considered is described in~\cite{lee_violeau:db3d:jhr2010} with a
block of fluid column of height $h = 0.55m$, width $w = 1.0m$ and length $l =
1.228m$. The container is $3.22m$ long. The block is released under gravity
with an acceleration of $-9.81 m/s^2$. Both schemes are simulated with a fixed
time step. For both schemes we use a CFL of 0.25. The speed of sound for the
EDAC case is set to $10 \sqrt{2gh}$, where $h$ is the height of the water
column. For the DTSPH, we use a time step of $\frac{0.25h}{\sqrt{2gh}}$, and
use $\beta=10, \epsilon=10^{-3}$. The particle spacing, $\Delta x=0.02$,
leading to around 240000 particles in the simulation.
%
%
\begin{figure}[!h]
  \centering
  \begin{subfigure}{0.4\linewidth}
    \includegraphics[width=1.0\linewidth]{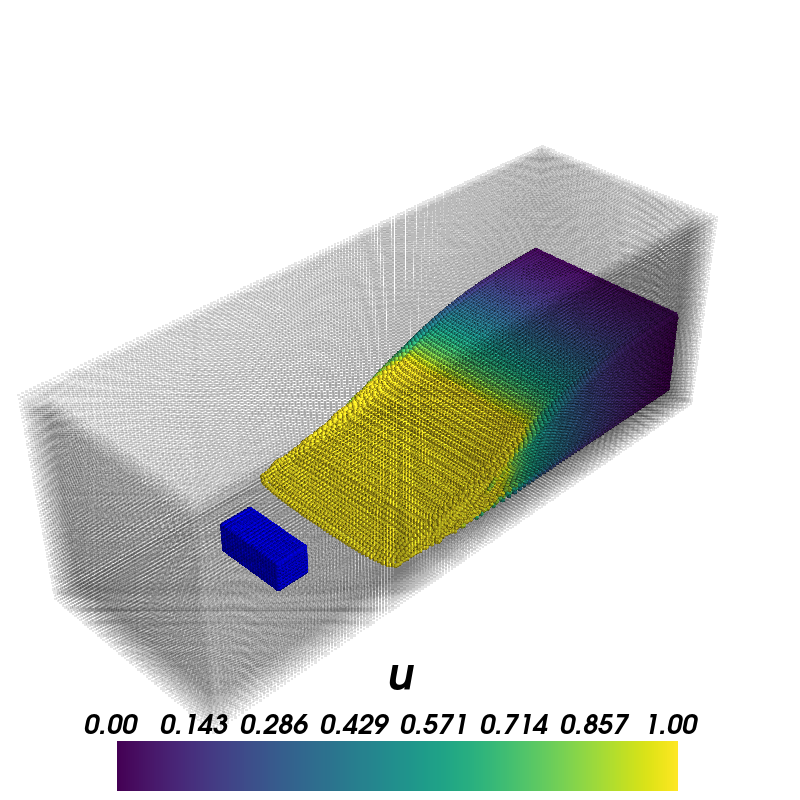}
  \end{subfigure}
  \begin{subfigure}{0.4\linewidth}
    \includegraphics[width=1.0\linewidth]{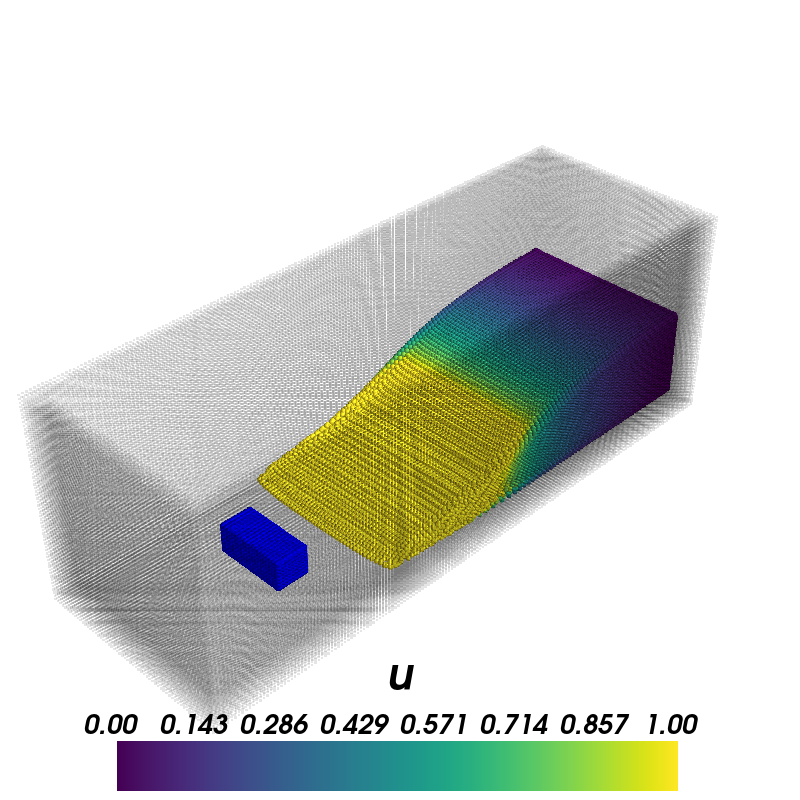}
  \end{subfigure}

  \begin{subfigure}{0.4\linewidth}
    \includegraphics[width=1.0\linewidth]{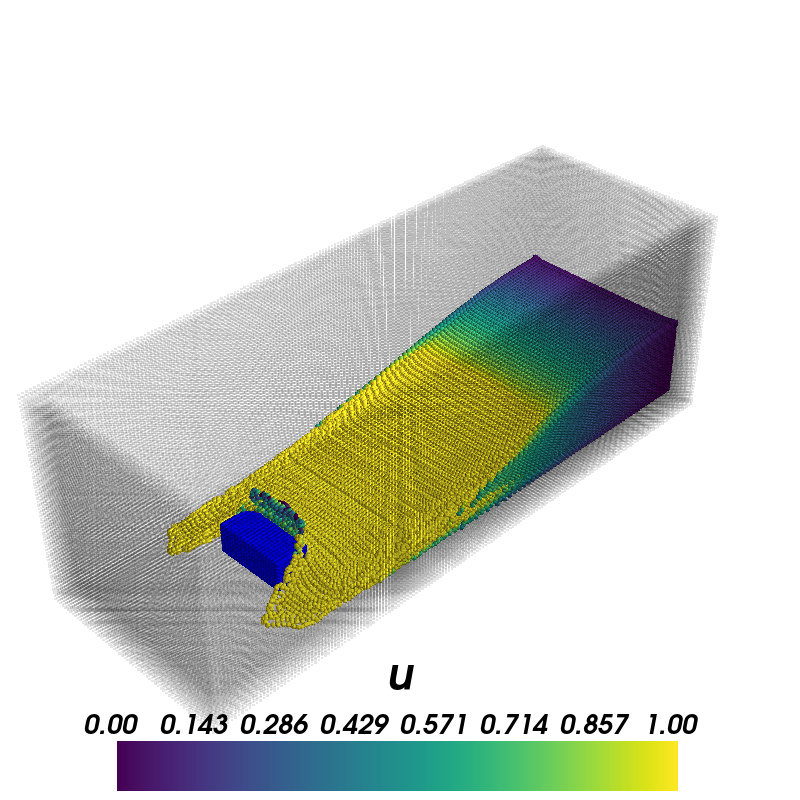}
  \end{subfigure}
  \begin{subfigure}{0.4\linewidth}
    \includegraphics[width=1.0\linewidth]{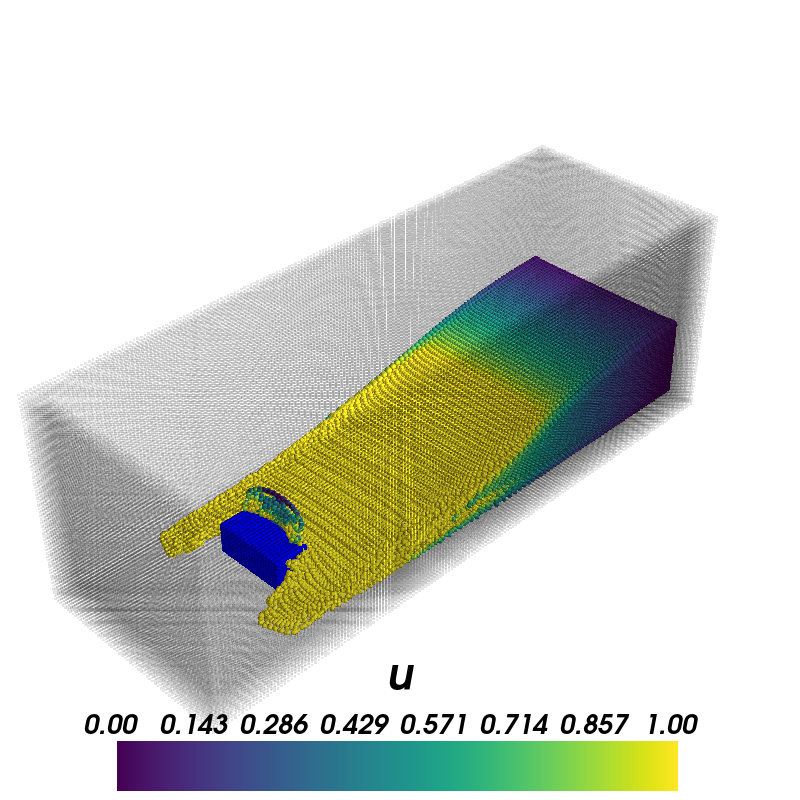}
  \end{subfigure}

  \begin{subfigure}{0.4\linewidth}
    \includegraphics[width=1.0\linewidth]{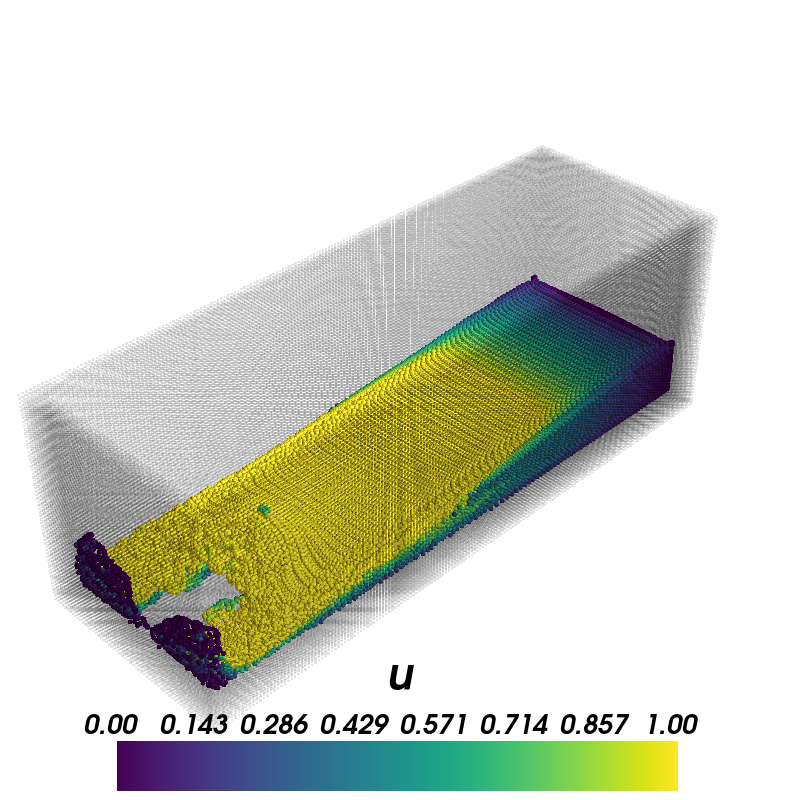}
    \caption{EDAC scheme.}
  \end{subfigure}
  \begin{subfigure}{0.4\linewidth}
    \includegraphics[width=1.0\linewidth]{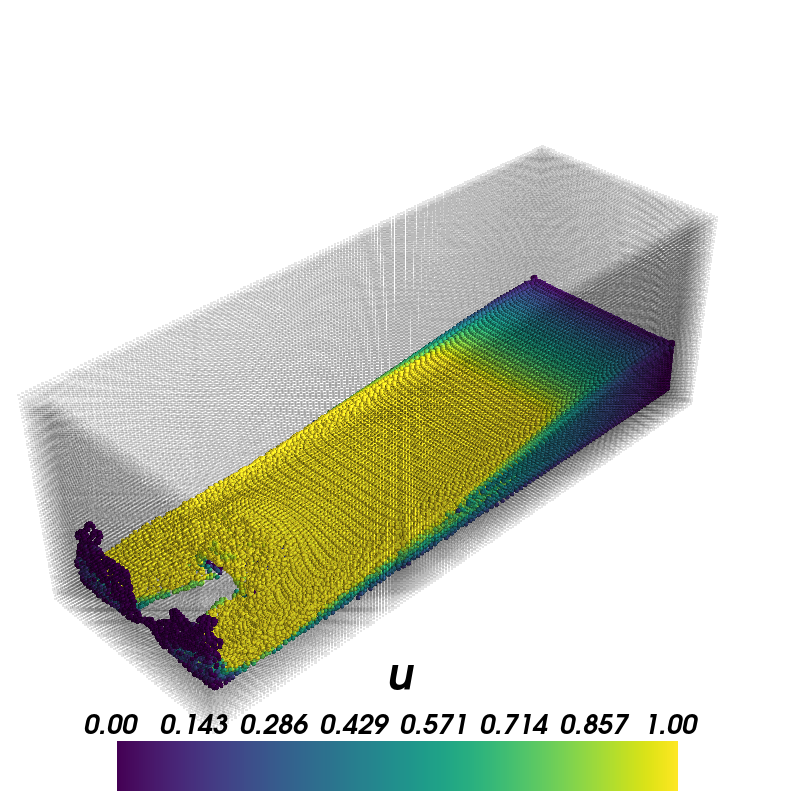}
    \caption{DTSPH scheme.}
  \end{subfigure}
  \caption{Dam-break 3D particle distribution at various times, first row is
    at $t = 0.4$~secs second row is at $t = 0.6$~secs and third row is at $t =
    1.0$~secs. The left column is EDAC scheme and right column is the DTSPH
    scheme.}
\label{fig:db3d:particle_plots_50}
\end{figure}

Fig.~\ref{fig:db3d:particle_plots_50} shows the particle distribution at
various times for both the schemes. This indicate that the new scheme produces
good results. Table~\ref{table:db3d:compare:times} shows the time taken for
the different 3D dam break simulations. Depending on the tolerance chosen, we
are able to obtain between a 1.7 to 7.16 fold improvement in performance as
compared to the EDAC scheme. We note that as the tolerance $\epsilon$ is
reduced, the DTSPH requires more iterations in pseudo-time in order to attain
convergence and this reduces the performance. However even with lower
tolerance values we are able to obtain very good results. In the next section
we demonstrate the performance achievable with the new scheme for different
problems.

\begin{table}[!htb]
\centering
\begin{tabular}{lrr}
\toprule
Scheme & $\epsilon$ &  CPU time (secs) \\
\midrule
 DTSPH &      0.001 &           563.19 \\
 DTSPH &     0.0001 &          2404.97 \\
  EDAC &         NA &          4035.25 \\
\bottomrule
\end{tabular}

\caption{CPU time time taken for different simulations of the 3D dam break
  problem.}\label{table:db3d:compare:times}
\end{table}

\subsection{Performance}
\label{dtsph:performance}

The performance of DTSPH is compared with that of other schemes for various
problems.  In Table~\ref{table:compare:times} we list the different problems
and the speedup obtained by using the new scheme.

As we have seen before, for the dam-break problem in three dimensions it can
be seen that DTSPH (with a tolerance of $10^{-3}$) can be up to 7.16 times
faster than the standard EDAC scheme. When a very low tolerance is used, the
scheme is about 1.7 times faster this is only to be expected as lower
tolerances require many more iterations to obtain a converged pressure. For
the two-dimensional dam break cases, the new scheme is about 3.5 times faster
than the EDAC or the $\delta$-SPH. For the unsteady cavity problem at $Re=100$
with a $150\times 150$ grid of particles, we get up to a 7 times performance
improvement. This is quite significant since in these cases we have compared
these cases where the effective Mach number is 0.1. Given this, the primary
advantage with the DTSPH is that it can take a time step that is 10 times
smaller. Hence in this case a speed-up of close to 7 suggests that it is an
efficient scheme.

\begin{table}[!htb]
\centering
\begin{tabular}{llr}
\toprule
Problem &  Scheme & Speed up\\
\midrule
Dam break 3D &   EDAC vs. DTSPH ($\epsilon = 10^{-3}$) &  7.16 \\
Dam break 3D &   EDAC vs. DTSPH ($\epsilon = 10^{-4}$) &  1.68 \\
Dam break 2D &   EDAC vs. DTSPH & 3.57 \\
  Dam break 2D &   EDAC vs. $\delta$-SPH & 3.66 \\
Cavity & TVF vs DTSPH & 6.87 \\
\bottomrule
\end{tabular}
\caption{Speed-up obtained for different simulations when using the DTSPH
  scheme.}
\label{table:compare:times}
\end{table}

The suite of benchmarks considered shows that the new scheme is robust,
simulates a variety of problems, and is as accurate as the EDAC scheme. In
addition it is very efficient and can be as much as seven times faster than
the EDAC scheme. Indeed, it is possible to improve the performance even more
by caching the values of the kernel and kernel gradients during the
pseudo-time iterations but we have not done this in the present work.


\section{Conclusions}
\label{sec:conclusions}

In this paper we propose a scheme called Dual-Time stepped SPH (DTSPH) that
employs a dual-time stepping approach for incompressible fluid flow
simulations. The method has been demonstrated with the EDAC
formulation~\cite{edac-sph:cf:2019}. There are a few recent developments in
the literature that employ a similar
approach~\cite{sph:acisph:cpc:2017,fatehi-2019} but these implementations are
not efficient despite the ability to use much smaller timesteps than the WCSPH
scheme. We find that by not moving the particles in pseudo-time we are able to
improve performance significantly without loss of accuracy as attested by our
simulations. We show that the scheme is robust and accurate. Through several
benchmarks in two and three dimensions we show that the scheme produces
results that are as accurate as the $\delta$-SPH
scheme~\cite{antuono-deltasph:cpc:2010} as well as the EDAC
scheme~\cite{edac-sph:cf:2019} while being up to seven times faster. The
method is matrix-free and may be implemented in the context of any explicit
SPH scheme. The DTSPH method does introduce a few new parameters in the form
of the term $\beta = \Delta t / \Delta \tau$ and the tolerance $\epsilon$. We
discuss how these parameters can be set based on rational considerations. For
a reasonable choice of parameters, the performance of the method is comparable
to that of incompressible SPH schemes. An open source implementation of the
new scheme is provided and the manuscript is fully reproducible. Given the
results presented with this method, a promising line of future work would be
to apply this to the case of solid mechanics problems where the timesteps are
very small. A method of this kind would provide significant computational
advantages. This would also be of considerable importance in the area of
fluid-structure-interaction which is an important emerging area of
investigation with the SPH method.

\section*{Acknowledgments}

We are grateful to the anonymous reviewers for their comments that have
improved the quality of this manuscript.

\afterpage{\clearpage}
\section*{References}
\bibliographystyle{model6-num-names}
\bibliography{references}

\begin{thebibliography}{39}
\providecommand{\natexlab}[1]{#1}
\providecommand{\url}[1]{\texttt{#1}}
\providecommand{\href}[2]{#2}
\providecommand{\path}[1]{#1}
\providecommand{\DOIprefix}{doi:}
\providecommand{\ArXivprefix}{arXiv:}
\providecommand{\URLprefix}{URL: }
\providecommand{\Pubmedprefix}{pmid:}
\providecommand{\doi}[1]{\href{http://dx.doi.org/#1}{\path{#1}}}
\providecommand{\Pubmed}[1]{\href{pmid:#1}{\path{#1}}}
\providecommand{\BIBand}{and}
\providecommand{\bibinfo}[2]{#2}
\ifx\xfnm\undefined \def\xfnm[#1]{\unskip,\space#1}\fi
\makeatletter\def\@biblabel#1{#1.}\makeatother
\bibitem[{Gingold and Monaghan(1977)}]{monaghan-gingold-stars-mnras-77}
\bibinfo{author}{Gingold\xfnm[ R.A.]}, \bibinfo{author}{Monaghan\xfnm[ J.J.]}.
\newblock \bibinfo{title}{Smoothed particle hydrodynamics: Theory and
  application to non-spherical stars}.
\newblock \emph{\bibinfo{journal}{Monthly Notices of the Royal Astronomical
  Society}}
  \bibinfo{year}{1977};\bibinfo{volume}{181}:\bibinfo{pages}{375--389}.
\bibitem[{Lucy(1977)}]{lucy77}
\bibinfo{author}{Lucy\xfnm[ L.B.]}.
\newblock \bibinfo{title}{{A numerical approach to testing the fission
  hypothesis}}.
\newblock \emph{\bibinfo{journal}{The Astronomical Journal}}
  \bibinfo{year}{1977};\bibinfo{volume}{82}(\bibinfo{number}{12}):\bibinfo{pages}{1013--1024}.
\bibitem[{Monaghan(1994)}]{sph:fsf:monaghan-jcp94}
\bibinfo{author}{Monaghan\xfnm[ J.J.]}.
\newblock \bibinfo{title}{Simulating free surface flows with {SPH}}.
\newblock \emph{\bibinfo{journal}{Journal of Computational Physics}}
  \bibinfo{year}{1994};\bibinfo{volume}{110}:\bibinfo{pages}{399--406}.
\bibitem[{Cummins and Rudman(1999)}]{sph:psph:cummins-rudman:jcp:1999}
\bibinfo{author}{Cummins\xfnm[ S.J.]}, \bibinfo{author}{Rudman\xfnm[ M.]}.
\newblock \bibinfo{title}{An {SPH} projection method}.
\newblock \emph{\bibinfo{journal}{Journal of Computational Physics}}
  \bibinfo{year}{1999};\bibinfo{volume}{152}:\bibinfo{pages}{584--607}.
\bibitem[{Shao and Lo(2003)}]{isph:shao:lo:awr:2003}
\bibinfo{author}{Shao\xfnm[ S.]}, \bibinfo{author}{Lo\xfnm[ E.Y.]}.
\newblock \bibinfo{title}{Incompressible {SPH} method for simulating newtonian
  and non-newtonian flows with a free surface}.
\newblock \emph{\bibinfo{journal}{Advances in Water Resources}}
  \bibinfo{year}{2003};\bibinfo{volume}{26}(\bibinfo{number}{7}):\bibinfo{pages}{787
  -- 800}.
\newblock \DOIprefix\doi{10.1016/S0309-1708(03)00030-7}.
\bibitem[{Gray et~al.(2001)Gray, Monaghan and
  Swift}]{sph-elastic-gray-jjm-cmame-2001}
\bibinfo{author}{Gray\xfnm[ J.]}, \bibinfo{author}{Monaghan\xfnm[ J.]},
  \bibinfo{author}{Swift\xfnm[ R.]}.
\newblock \bibinfo{title}{Sph elastic dynamics}.
\newblock \emph{\bibinfo{journal}{Computer Methods in Applied Mechanics and
  Engineering}}
  \bibinfo{year}{2001};\bibinfo{volume}{190}(\bibinfo{number}{49}):\bibinfo{pages}{6641--6662}.
\newblock \DOIprefix\doi{10.1016/S0045-7825(01)00254-7}.
\bibitem[{Rafiee and Thiagarajan(2009)}]{rafiee:fsi-2009}
\bibinfo{author}{Rafiee\xfnm[ A.]}, \bibinfo{author}{Thiagarajan\xfnm[ K.P.]}.
\newblock \bibinfo{title}{An {SPH} projection method for simulating
  fluid-hypoelastic structure interaction}.
\newblock \emph{\bibinfo{journal}{Computer Methods in Applied Mechanics and
  Engineering}}
  \bibinfo{year}{2009};\bibinfo{volume}{198}(\bibinfo{number}{33}):\bibinfo{pages}{2785--2795}.
\newblock \DOIprefix\doi{10.1016/j.cma.2009.04.001}.
\bibitem[{Khayyer et~al.(2018)Khayyer, Gotoh, Falahaty and
  Shimizu}]{khayyer-fsi-2018}
\bibinfo{author}{Khayyer\xfnm[ A.]}, \bibinfo{author}{Gotoh\xfnm[ H.]},
  \bibinfo{author}{Falahaty\xfnm[ H.]}, \bibinfo{author}{Shimizu\xfnm[ Y.]}.
\newblock \bibinfo{title}{An enhanced {ISPH}–{SPH} coupled method for
  simulation of incompressible fluid–elastic structure interactions}.
\newblock \emph{\bibinfo{journal}{Computer Physics Communications}}
  \bibinfo{year}{2018};\bibinfo{volume}{232}:\bibinfo{pages}{139--164}.
\newblock \DOIprefix\doi{10.1016/j.cpc.2018.05.012}.
\bibitem[{Adami et~al.(2013)Adami, Hu and Adams}]{Adami2013}
\bibinfo{author}{Adami\xfnm[ S.]}, \bibinfo{author}{Hu\xfnm[ X.]},
  \bibinfo{author}{Adams\xfnm[ N.]}.
\newblock \bibinfo{title}{{A transport-velocity formulation for smoothed
  particle hydrodynamics}}.
\newblock \emph{\bibinfo{journal}{Journal of Computational Physics}}
  \bibinfo{year}{2013};\bibinfo{volume}{241}:\bibinfo{pages}{292--307}.
\newblock \DOIprefix\doi{10.1016/j.jcp.2013.01.043}.
\bibitem[{Antuono et~al.(2010)Antuono, Colagrossi, Marrone and
  Molteni}]{antuono-deltasph:cpc:2010}
\bibinfo{author}{Antuono\xfnm[ M.]}, \bibinfo{author}{Colagrossi\xfnm[ A.]},
  \bibinfo{author}{Marrone\xfnm[ S.]}, \bibinfo{author}{Molteni\xfnm[ D.]}.
\newblock \bibinfo{title}{Free-surface flows solved by means of {SPH} schemes
  with numerical diffusive terms}.
\newblock \emph{\bibinfo{journal}{Computer Physics Communications}}
  \bibinfo{year}{2010};\bibinfo{volume}{181}(\bibinfo{number}{3}):\bibinfo{pages}{532
  -- 549}.
\newblock \DOIprefix\doi{10.1016/j.cpc.2009.11.002}.
\bibitem[{{Marrone} et~al.(2011){Marrone}, {Antuono}, {Colagrossi},
  {Colicchio}, {Le Touz{\'e}} and {Graziani}}]{marrone-deltasph:cmame:2011}
\bibinfo{author}{{Marrone}\xfnm[ S.]}, \bibinfo{author}{{Antuono}\xfnm[ M.]},
  \bibinfo{author}{{Colagrossi}\xfnm[ A.]}, \bibinfo{author}{{Colicchio}\xfnm[
  G.]}, \bibinfo{author}{{Le Touz{\'e}}\xfnm[ D.]},
  \bibinfo{author}{{Graziani}\xfnm[ G.]}.
\newblock \bibinfo{title}{{$\delta$-SPH} model for simulating violent impact
  flows}.
\newblock \emph{\bibinfo{journal}{Computer Methods in Applied Mechanics and
  Engineering}}
  \bibinfo{year}{2011};\bibinfo{volume}{200}:\bibinfo{pages}{1526--1542}.
\newblock \DOIprefix\doi{10.1016/j.cma.2010.12.016}.
\bibitem[{Ramachandran and Puri(2019)}]{edac-sph:cf:2019}
\bibinfo{author}{Ramachandran\xfnm[ P.]}, \bibinfo{author}{Puri\xfnm[ K.]}.
\newblock \bibinfo{title}{Entropically damped artificial compressibility for
  {SPH}}.
\newblock \emph{\bibinfo{journal}{Computers and Fluids}}
  \bibinfo{year}{2019};\bibinfo{volume}{179}(\bibinfo{number}{30}):\bibinfo{pages}{579--594}.
\newblock \DOIprefix\doi{10.1016/j.compfluid.2018.11.023}.
\bibitem[{Hu and Adams(2007)}]{isph:hu-adams:jcp:2007}
\bibinfo{author}{Hu\xfnm[ X.]}, \bibinfo{author}{Adams\xfnm[ N.]}.
\newblock \bibinfo{title}{{An incompressible multi-phase SPH method}}.
\newblock \emph{\bibinfo{journal}{Journal of Computational Physics}}
  \bibinfo{year}{2007};\bibinfo{volume}{227}(\bibinfo{number}{1}):\bibinfo{pages}{264--278}.
\newblock \DOIprefix\doi{10.1016/j.jcp.2007.07.013}.
\bibitem[{Solenthaler and Pajarola(2009)}]{pcisph:acm_tog:2009}
\bibinfo{author}{Solenthaler\xfnm[ B.]}, \bibinfo{author}{Pajarola\xfnm[ R.]}.
\newblock \bibinfo{title}{Predictive-corrective incompressible {SPH}}.
\newblock \emph{\bibinfo{journal}{ACM Transactions on Graphics}}
  \bibinfo{year}{2009};\bibinfo{volume}{28}(\bibinfo{number}{3}):\bibinfo{pages}{40:1--40:6}.
\newblock \DOIprefix\doi{10.1145/1531326.1531346}.
\bibitem[{Ihmsen et~al.(2014)Ihmsen, Cornelis, Solenthaler, Horvath and
  Teschner}]{iisph:ihmsen:tvcg-2014}
\bibinfo{author}{Ihmsen\xfnm[ M.]}, \bibinfo{author}{Cornelis\xfnm[ J.]},
  \bibinfo{author}{Solenthaler\xfnm[ B.]}, \bibinfo{author}{Horvath\xfnm[ C.]},
  \bibinfo{author}{Teschner\xfnm[ M.]}.
\newblock \bibinfo{title}{Implicit incompressible {SPH}}.
\newblock \emph{\bibinfo{journal}{{IEEE} Trans Vis Comput Graph}}
  \bibinfo{year}{2014};\bibinfo{volume}{20}(\bibinfo{number}{3}):\bibinfo{pages}{426--435}.
\newblock \DOIprefix\doi{10.1109/TVCG.2013.105}.
\bibitem[{Muta et~al.(2020)Muta, Ramachandran and Negi}]{sisph}
\bibinfo{author}{Muta\xfnm[ A.]}, \bibinfo{author}{Ramachandran\xfnm[ P.]},
  \bibinfo{author}{Negi\xfnm[ P.]}.
\newblock \bibinfo{title}{An efficient, open source, iterative isph scheme}.
\newblock \emph{\bibinfo{journal}{Computer Physics Communications}}
  \bibinfo{year}{2020};\bibinfo{volume}{255}:\bibinfo{pages}{107283}.
\newblock \DOIprefix\doi{10.1016/j.cpc.2020.107283}.
\bibitem[{Rouzbahani and Hejranfar(2017)}]{sph:acisph:cpc:2017}
\bibinfo{author}{Rouzbahani\xfnm[ F.]}, \bibinfo{author}{Hejranfar\xfnm[ K.]}.
\newblock \bibinfo{title}{A truly incompressible smoothed particle
  hydrodynamics based on artificial compressibility method}.
\newblock \emph{\bibinfo{journal}{Computer Physics Communications}}
  \bibinfo{year}{2017};\bibinfo{volume}{210}:\bibinfo{pages}{10 -- 28}.
\newblock \DOIprefix\doi{10.1016/j.cpc.2016.09.008}.
\bibitem[{Chorin(1967)}]{art-compr-chorin-jcp-67}
\bibinfo{author}{Chorin\xfnm[ A.J.]}.
\newblock \bibinfo{title}{A numerical method for solving incompressible viscous
  flow problems}.
\newblock \emph{\bibinfo{journal}{Journal of Computational Physics}}
  \bibinfo{year}{1967};\bibinfo{volume}{2}(\bibinfo{number}{1}):\bibinfo{pages}{12
  -- 26}.
\newblock \DOIprefix\doi{10.1016/0021-9991(67)90037-X}.
\bibitem[{Fatehi et~al.(2019)Fatehi, Rahmat, Tofighi, Yildiz and
  Shadloo}]{fatehi-2019}
\bibinfo{author}{Fatehi\xfnm[ R.]}, \bibinfo{author}{Rahmat\xfnm[ A.]},
  \bibinfo{author}{Tofighi\xfnm[ N.]}, \bibinfo{author}{Yildiz\xfnm[ M.]},
  \bibinfo{author}{Shadloo\xfnm[ M.S.]}.
\newblock \bibinfo{title}{Density-based smoothed particle hydrodynamics methods
  for incompressible flows}.
\newblock \emph{\bibinfo{journal}{Computers \& Fluids}}
  \bibinfo{year}{2019};\bibinfo{volume}{185}:\bibinfo{pages}{22--33}.
\newblock \DOIprefix\doi{10.1016/j.compfluid.2019.02.018}.
\bibitem[{Zhang et~al.(2020)Zhang, Rezavand and Hu}]{zhang-dual-time2020}
\bibinfo{author}{Zhang\xfnm[ C.]}, \bibinfo{author}{Rezavand\xfnm[ M.]},
  \bibinfo{author}{Hu\xfnm[ X.]}.
\newblock \bibinfo{title}{Dual-criteria time stepping for weakly compressible
  smoothed particle hydrodynamics}.
\newblock \emph{\bibinfo{journal}{Journal of Computational Physics}}
  \bibinfo{year}{2020};\bibinfo{volume}{404}:\bibinfo{pages}{109135}.
\newblock \DOIprefix\doi{10.1016/j.jcp.2019.109135}.
\bibitem[{Ramachandran et~al.(2020)Ramachandran, Bhosale, Puri, Negi, Muta,
  Adepu, Menon, Govind, Sanka, Sebastian, Sen, Kaushik, Kumar, Kurapati, Patil,
  Tavker, Pandey, Kaushik, Dutt and Agarwal}]{pysph2020}
\bibinfo{author}{Ramachandran\xfnm[ P.]}, \bibinfo{author}{Bhosale\xfnm[ A.]},
  \bibinfo{author}{Puri\xfnm[ K.]}, \bibinfo{author}{Negi\xfnm[ P.]},
  \bibinfo{author}{Muta\xfnm[ A.]}, \bibinfo{author}{Adepu\xfnm[ D.]},
  \bibinfo{author}{Menon\xfnm[ D.]}, \bibinfo{author}{Govind\xfnm[ R.]},
  \bibinfo{author}{Sanka\xfnm[ S.]}, \bibinfo{author}{Sebastian\xfnm[ A.S.]},
  \bibinfo{author}{Sen\xfnm[ A.]}, \bibinfo{author}{Kaushik\xfnm[ R.]},
  \bibinfo{author}{Kumar\xfnm[ A.]}, \bibinfo{author}{Kurapati\xfnm[ V.]},
  \bibinfo{author}{Patil\xfnm[ M.]}, \bibinfo{author}{Tavker\xfnm[ D.]},
  \bibinfo{author}{Pandey\xfnm[ P.]}, \bibinfo{author}{Kaushik\xfnm[ C.]},
  \bibinfo{author}{Dutt\xfnm[ A.]}, \bibinfo{author}{Agarwal\xfnm[ A.]}.
\newblock \bibinfo{title}{{PySPH}: a {Python-based} framework for smoothed
  particle hydrodynamics}.
\newblock \emph{\bibinfo{journal}{arXiv preprint arXiv:190904504}}
  \bibinfo{year}{2020};\URLprefix \url{https://arxiv.org/abs/1909.04504}.
\bibitem[{Ramachandran(2016)}]{PR:pysph:scipy16}
\bibinfo{author}{Ramachandran\xfnm[ P.]}.
\newblock \bibinfo{title}{{PySPH}: a reproducible and high-performance
  framework for smoothed particle hydrodynamics}.
\newblock In: \bibinfo{editor}{Benthall\xfnm[ S.]},
  \bibinfo{editor}{Rostrup\xfnm[ S.]}, eds.
  \emph{\bibinfo{booktitle}{{P}roceedings of the 15th {P}ython in {S}cience
  {C}onference}}. \bibinfo{year}{2016}:\unskip \bibinfo{pages}{127 -- 135}.
\newblock \DOIprefix\doi{10.25080/Majora-629e541a-011}.
\bibitem[{Ramachandran(2018)}]{pr:automan:2018}
\bibinfo{author}{Ramachandran\xfnm[ P.]}.
\newblock \bibinfo{title}{automan: A python-based automation framework for
  numerical computing}.
\newblock \emph{\bibinfo{journal}{Computing in Science \& Engineering}}
  \bibinfo{year}{2018};\bibinfo{volume}{20}(\bibinfo{number}{5}):\bibinfo{pages}{81--97}.
\newblock \DOIprefix\doi{10.1109/MCSE.2018.05329818}.
\bibitem[{Monaghan(2005)}]{monaghan-review:2005}
\bibinfo{author}{Monaghan\xfnm[ J.J.]}.
\newblock \bibinfo{title}{{Smoothed Particle Hydrodynamics}}.
\newblock \emph{\bibinfo{journal}{{Reports on Progress in Physics}}}
  \bibinfo{year}{2005};\bibinfo{volume}{68}:\bibinfo{pages}{1703--1759}.
\bibitem[{Morris et~al.(1997)Morris, Fox and Zhu}]{morris-lowRe-97}
\bibinfo{author}{Morris\xfnm[ J.P.]}, \bibinfo{author}{Fox\xfnm[ P.J.]},
  \bibinfo{author}{Zhu\xfnm[ Y.]}.
\newblock \bibinfo{title}{Modeling low reynolds number incompressible flows
  using {SPH}}.
\newblock \emph{\bibinfo{journal}{Journal of Computational Physics}}
  \bibinfo{year}{1997};\bibinfo{volume}{136}(\bibinfo{number}{1}):\bibinfo{pages}{214--226}.
\newblock \DOIprefix\doi{10.1006/jcph.1997.5776}.
\bibitem[{Ferrari et~al.(2009)Ferrari, Dumbser, Toro and
  Armanini}]{ferrari-riemann-sph-2009}
\bibinfo{author}{Ferrari\xfnm[ A.]}, \bibinfo{author}{Dumbser\xfnm[ M.]},
  \bibinfo{author}{Toro\xfnm[ E.F.]}, \bibinfo{author}{Armanini\xfnm[ A.]}.
\newblock \bibinfo{title}{A new {{3D}} parallel {{SPH}} scheme for free surface
  flows}.
\newblock \emph{\bibinfo{journal}{Computers \& Fluids}}
  \bibinfo{year}{2009};\bibinfo{volume}{38}(\bibinfo{number}{6}):\bibinfo{pages}{1203--1217}.
\newblock \DOIprefix\doi{10.1016/j.compfluid.2008.11.012}.
\bibitem[{Adami et~al.(2012)Adami, Hu and Adams}]{Adami2012}
\bibinfo{author}{Adami\xfnm[ S.]}, \bibinfo{author}{Hu\xfnm[ X.]},
  \bibinfo{author}{Adams\xfnm[ N.]}.
\newblock \bibinfo{title}{{A generalized wall boundary condition for smoothed
  particle hydrodynamics}}.
\newblock \emph{\bibinfo{journal}{Journal of Computational Physics}}
  \bibinfo{year}{2012};\bibinfo{volume}{231}(\bibinfo{number}{21}):\bibinfo{pages}{7057--7075}.
\newblock \DOIprefix\doi{10.1016/j.jcp.2012.05.005}.
\bibitem[{Hughes and Graham(2010)}]{hughes-graham:compare-wcsph:jhr:2010}
\bibinfo{author}{Hughes\xfnm[ J.]}, \bibinfo{author}{Graham\xfnm[ D.]}.
\newblock \bibinfo{title}{Comparison of incompressible and weakly-compressible
  {SPH} models for free-surface water flows}.
\newblock \emph{\bibinfo{journal}{Journal of Hydraulic Research}}
  \bibinfo{year}{2010};\bibinfo{volume}{48}:\bibinfo{pages}{105--117}.
\bibitem[{Lind et~al.(2012)Lind, Xu, Stansby and
  Rogers}]{diff_smoothing_sph:lind:jcp:2009}
\bibinfo{author}{Lind\xfnm[ S.]}, \bibinfo{author}{Xu\xfnm[ R.]},
  \bibinfo{author}{Stansby\xfnm[ P.]}, \bibinfo{author}{Rogers\xfnm[ B.]}.
\newblock \bibinfo{title}{Incompressible smoothed particle hydrodynamics for
  free-surface flows: A generalised diffusion-based algorithm for stability and
  validations for impulsive flows and propagating waves}.
\newblock \emph{\bibinfo{journal}{Journal of Computational Physics}}
  \bibinfo{year}{2012};\bibinfo{volume}{231}(\bibinfo{number}{4}):\bibinfo{pages}{1499
  -- 1523}.
\newblock \DOIprefix\doi{10.1016/j.jcp.2011.10.027}.
\bibitem[{Ghasemi~V. et~al.(2013)Ghasemi~V., Firoozabadi and
  Mahdinia}]{ghasemi-2013}
\bibinfo{author}{Ghasemi~V.\xfnm[ A.]}, \bibinfo{author}{Firoozabadi\xfnm[
  B.]}, \bibinfo{author}{Mahdinia\xfnm[ M.]}.
\newblock \bibinfo{title}{{{2D}} numerical simulation of density currents using
  the {{SPH}} projection method}.
\newblock \emph{\bibinfo{journal}{European Journal of Mechanics - B/Fluids}}
  \bibinfo{year}{2013};\bibinfo{volume}{38}:\bibinfo{pages}{38--46}.
\newblock \DOIprefix\doi{10.1016/j.euromechflu.2012.10.004}.
\bibitem[{Lee et~al.(2008)Lee, Moulinec, Xu, Violeau, Laurence and
  Stansby}]{lee-comparison-2008}
\bibinfo{author}{Lee\xfnm[ E.S.]}, \bibinfo{author}{Moulinec\xfnm[ C.]},
  \bibinfo{author}{Xu\xfnm[ R.]}, \bibinfo{author}{Violeau\xfnm[ D.]},
  \bibinfo{author}{Laurence\xfnm[ D.]}, \bibinfo{author}{Stansby\xfnm[ P.]}.
\newblock \bibinfo{title}{Comparisons of weakly compressible and truly
  incompressible algorithms for the {{SPH}} mesh free particle method}.
\newblock \emph{\bibinfo{journal}{Journal of Computational Physics}}
  \bibinfo{year}{2008};\bibinfo{volume}{227}(\bibinfo{number}{18}):\bibinfo{pages}{8417--8436}.
\newblock \DOIprefix\doi{10.1016/j.jcp.2008.06.005}.
\bibitem[{Ghia et~al.(1982)Ghia, Ghia and Shin}]{ldc:ghia-1982}
\bibinfo{author}{Ghia\xfnm[ U.]}, \bibinfo{author}{Ghia\xfnm[ K.N.]},
  \bibinfo{author}{Shin\xfnm[ C.T.]}.
\newblock \bibinfo{title}{{High-Re} solutions for incompressible flow using the
  {Navier-Stokes} equations and a multigrid method}.
\newblock \emph{\bibinfo{journal}{Journal of Computational Physics}}
  \bibinfo{year}{1982};\bibinfo{volume}{48}:\bibinfo{pages}{387--411}.
\bibitem[{Colagrossi(2005)}]{colagrossi-phdthesis:2005}
\bibinfo{author}{Colagrossi\xfnm[ A.]}.
\newblock \bibinfo{title}{A meshless lagrangian method for free-surface and
  interface flows with fragmentation}.
\newblock \emph{\bibinfo{journal}{These, Universita di Roma}}
  \bibinfo{year}{2005};\URLprefix \url{http://hdl.handle.net/10805/688}.
\bibitem[{Khayyer and Gotoh(2013)}]{khayyer-2013}
\bibinfo{author}{Khayyer\xfnm[ A.]}, \bibinfo{author}{Gotoh\xfnm[ H.]}.
\newblock \bibinfo{title}{Enhancement of performance and stability of {{MPS}}
  mesh-free particle method for multiphase flows characterized by high density
  ratios}.
\newblock \emph{\bibinfo{journal}{Journal of Computational Physics}}
  \bibinfo{year}{2013};\bibinfo{volume}{242}:\bibinfo{pages}{211--233}.
\newblock \DOIprefix\doi{10.1016/j.jcp.2013.02.002}.
\bibitem[{Sun et~al.(2017)Sun, Colagrossi, Marrone and Zhang}]{sun-deltap-2017}
\bibinfo{author}{Sun\xfnm[ P.]}, \bibinfo{author}{Colagrossi\xfnm[ A.]},
  \bibinfo{author}{Marrone\xfnm[ S.]}, \bibinfo{author}{Zhang\xfnm[ A.]}.
\newblock \bibinfo{title}{The {$\delta^{+}$}-{{SPH}} model: {{Simple}}
  procedures for a further improvement of the {{SPH}} scheme}.
\newblock \emph{\bibinfo{journal}{Computer Methods in Applied Mechanics and
  Engineering}}
  \bibinfo{year}{2017};\bibinfo{volume}{315}:\bibinfo{pages}{25--49}.
\newblock \DOIprefix\doi{10.1016/j.cma.2016.10.028}.
\bibitem[{Lind et~al.(2016)Lind, Stansby and Rogers}]{lind-2016}
\bibinfo{author}{Lind\xfnm[ S.]}, \bibinfo{author}{Stansby\xfnm[ P.]},
  \bibinfo{author}{Rogers\xfnm[ B.]}.
\newblock \bibinfo{title}{Incompressible\textendash compressible flows with a
  transient discontinuous interface using smoothed particle hydrodynamics
  ({{SPH}})}.
\newblock \emph{\bibinfo{journal}{Journal of Computational Physics}}
  \bibinfo{year}{2016};\bibinfo{volume}{309}:\bibinfo{pages}{129--147}.
\newblock \DOIprefix\doi{10.1016/j.jcp.2015.12.005}.
\bibitem[{Rezavand et~al.(2018)Rezavand, {Taeibi-Rahni} and
  Rauch}]{rezavand-2018a}
\bibinfo{author}{Rezavand\xfnm[ M.]}, \bibinfo{author}{{Taeibi-Rahni}\xfnm[
  M.]}, \bibinfo{author}{Rauch\xfnm[ W.]}.
\newblock \bibinfo{title}{An {{ISPH}} scheme for numerical simulation of
  multiphase flows with complex interfaces and high density ratios}.
\newblock \emph{\bibinfo{journal}{Computers \& Mathematics with Applications}}
  \bibinfo{year}{2018};\bibinfo{volume}{75}(\bibinfo{number}{8}):\bibinfo{pages}{2658--2677}.
\newblock \DOIprefix\doi{10.1016/j.camwa.2017.12.034}.
\bibitem[{Lee et~al.(2010)Lee, Violeau, Issa and
  Ploix}]{lee_violeau:db3d:jhr2010}
\bibinfo{author}{Lee\xfnm[ E.S.]}, \bibinfo{author}{Violeau\xfnm[ D.]},
  \bibinfo{author}{Issa\xfnm[ R.]}, \bibinfo{author}{Ploix\xfnm[ S.]}.
\newblock \bibinfo{title}{Application of weakly compressible and truly
  incompressible {SPH} to 3-d water collapse in waterworks}.
\newblock \emph{\bibinfo{journal}{Journal of Hydraulic Research}}
  \bibinfo{year}{2010};\bibinfo{volume}{48}:\bibinfo{pages}{50--60}.
\newblock \DOIprefix\doi{10.1080/00221686.2010.9641245}.
\bibitem[{Koshizuka and Oka(1996)}]{koshizuka_oka_mps:nse:1996}
\bibinfo{author}{Koshizuka\xfnm[ S.]}, \bibinfo{author}{Oka\xfnm[ Y.]}.
\newblock \bibinfo{title}{Moving-particle semi-implicit method for
  fragmentation of incompressible fluid}.
\newblock \emph{\bibinfo{journal}{Nuclear Science and Engineering}}
  \bibinfo{year}{1996};\bibinfo{volume}{123}:\bibinfo{pages}{421--434}.

\end{thebibliography}

\section*{Appendix}

This section provides a derivation of the perturbation velocity that is given
in equation~\eqref{eq:vtilde}.

Using trapezoidal rule for integration, the displacement of the particle in
pseudo time from the initial state $(\ten{r}^{k=0}, \ten{V}^{k=0}, 0)$ to the
current pseudo time state  $(\ten{r}^k, \ten{V}^{k+1}, \Delta t)$ when $k
\rightarrow \infty$ is given by,
\begin{equation}
  \label{eq:pos-vel-trapz-rule}
  \ten{r}^{k+1} - \ten{r}^{n} = \Delta t \frac{(\ten{V}^{k+1}
  + \ten{V}^{n})}{2},
\end{equation}
similarly, using the trapezoidal rule of integration for the displacement
between states in pseudo time is given by,
\begin{equation}
  \label{eq:pos-update-simple}
  \ten{r}^{k+1} = \ten{r}^{k} + \frac{\Delta \tau}{2} (\ten{\tilde{V}}^{k}
  + \ten{\tilde{V}}^{k+1}).
\end{equation}
Expanding $\ten{r}^k$ in terms of $\ten{r}^{0}$ (i.e.\ $k=0$)is,
\begin{equation}
  \label{eq:pos-update-expand}
  \ten{r}^{k+1} = \ten{r}^{0} +
  \frac{\Delta \tau}{2} \ten{\tilde{V}}^{0}
  + \Delta \tau \sum_{j=1}^{k}\ten{\tilde{V}}^{j} + \frac{\Delta \tau}{2}
  \ten{\tilde{V}}^{k+1},
\end{equation}
where, the position before pseudo time iteration $\ten{r}^{0}$ is given
by~\eqref{eq:r0}.

Substitute the above equation~\eqref{eq:pos-update-expand} into
the equation~\eqref{eq:pos-vel-trapz-rule} we get,
\begin{equation}
  \label{eq:sub-pos-update}
  \ten{r}^0 - \ten{r}^n + \frac{\Delta \tau}{2}(\ten{\tilde{V}}^{0}
  + \ten{\tilde{V}}^{k+1}) + \Delta \tau \sum_{j=1}^{k} \ten{\tilde{V}}^{j}
  = \Delta t \frac{(\ten{V}^{k+1} + \ten{V}^{n})}{2}.
\end{equation}
Rearranging terms to get $\ten{V}^{k+1}$ as,
\begin{equation}
  \label{eq:vkupdate}
  \ten{V}^{k+1} =
  \frac{2}{\Delta t} \left( \ten{r}^0 - \ten{r}^n \right) +
  \frac{\Delta \tau}{\Delta t}(\ten{\tilde{V}}^{0}
  + \ten{\tilde{V}}^{k+1}) + \frac{2\Delta \tau}{\Delta t}
  \sum_{j=1}^{k} \ten{\tilde{V}}^{j}
  - \ten{V}^{n}.
\end{equation}
Similarly $\ten{V}^{k}$ is written as,
\begin{equation}
  \label{eq:vkupdate-prev}
  \ten{V}^{k} =
  \frac{2}{\Delta t} \left( \ten{r}^0 - \ten{r}^n \right) +
  \frac{\Delta \tau}{\Delta t}(\ten{\tilde{V}}^{0}
  + \ten{\tilde{V}}^{k}) + \frac{2\Delta \tau}{\Delta t}
  \sum_{j=1}^{k-1} \ten{\tilde{V}}^{j}
  - \ten{V}^{n}.
\end{equation}
Subtract~\eqref{eq:vkupdate-prev} from~\eqref{eq:vkupdate},
\begin{equation}
  \label{eq:sub-vkupdate}
  \ten{V}^{k+1} - \ten{V}^{k}=
  \frac{\Delta \tau}{\Delta t}(\ten{\tilde{V}}^{k+1} + \ten{\tilde{V}}^{k}).
\end{equation}
By substituting eq.~(\ref{eq:vkp1}) we get,
\begin{equation}
  \label{eq:vtilde-proof}
  \ten{\tilde{V}}^{k}
  = \frac{\Delta t}{\Delta \tau}(\ten{V}^{k+1} - \ten{V}^{k})
  = \Delta t {\left( \frac{d \ten{V}}{d \tau} \right)}^{k+1/2}.
\end{equation}
Note that in the limit $k \rightarrow \infty$, $\ten{\tilde{V}}^{k+1}$ goes to
zero, and $\lim_{k \rightarrow \infty} \ten{V}^{k+1} = \ten{V}^{n+1}$.

\end{document}